\documentclass[10pt]{iopart}

\usepackage{graphicx}
\usepackage[colorlinks=true,citecolor=blue,urlcolor=blue,linkcolor=blue]{hyperref}
\hypersetup{colorlinks=true,linktoc=all,linkcolor=blue,breaklinks=true,citecolor=blue,urlcolor=blue}
\usepackage{bbold}
\usepackage{xcolor}
\usepackage{units}
\usepackage{booktabs}
\usepackage{array}
\usepackage[normalem]{ulem}

\usepackage{mathrsfs}
\usepackage{bm}
\usepackage{mathbbol}
\usepackage{epsfig}
\usepackage{url}
\usepackage{stackrel}
\usepackage{amssymb}

\newcommand{\bea}{\begin{eqnarray}}
\newcommand{\eea}{\end{eqnarray}}
\newcommand{\beq}{\begin{equation}}
\newcommand{\eeq}{\end{equation}}
\newcommand{\rmee}{{\rm e}}

\newcommand{\nB}{n_{\rm B}}

\newcommand{\kBT}{k_{\rm B}T}

\newcommand{\eqref}[1]{(\ref{#1})}

\begin{document}

\title[The qutrit as a heat diode and circulator]{
The qutrit as a heat diode and circulator}

\author{Israel D\'iaz$^1$ and Rafael S\'anchez$^{2}$}
\address{$^1$ Universidad Aut\'onoma de Madrid, 28049 Madrid, Spain}
\address{$^2$ Departamento de F\'isica Te\'orica de la Materia Condensada, Condensed Matter Physics Center (IFIMAC) and Insituto Nicol\'as Cabrera, Universidad Aut\'onoma de Madrid, 28049 Madrid, Spain}
\vspace{10pt}

\begin{abstract}
We investigate the heat transport properties of a three-level system coupled to three thermal baths, assuming a model based on superconducting circuit implementations. The system-bath coupling is mediated by resonators which serve as frequency filters for the different qutrit transitions. Taking into account the finite quality factors of the resonators, we find thermal rectification and circulation effects not expected in configurations with perfectly-filtered couplings. Heat leakage in off-resonant transitions can be exploited to make the system work as an ideal diode where heat flows in the same direction between two baths irrespective of the sign of the temperature difference, as well as a perfect heat circulator whose state is phase-reversible.
\end{abstract}

%
%
%
%
%

\section{Introduction}
\label{sec:intro}

The problem of heat transport and thermodynamic operations related to the dynamics of few level quantum systems in contact to thermal baths has attracted interest for decades~\cite{scovil_three_1959,geusic_quantum_1967,kosloff:1984,mu_oneatom_1992} which has recently been reactivated in the context of quantum thermodynamics~\cite{kosloff_quantum_2014,mitchison:2020,thermo:book}. Questions related to the limitations of refrigeration~\cite{palao_quantum_2001,linden_how_2010,
levy_quantum_2012,levy:2012pre,correa_performance_2013,brunner_entanglement_2014,
mohanta_universal_2021}, quantum heat engines~\cite{scully_quantum_2011,brunner_virtual_2012,silva_small_2015,
mitchison:2016,hewgill_three_2020,naseem_two_2020,bhandari_minimal_2021,
poulsen_nonmarkovian_2021}, the role of entanglement~\cite{blok_quantum_2021,brask:2015njp,nesterov_proposal_2021} and fluctuations~\cite{li_quantum_2017,mohanta_universal_2021,kalaee_violating_2021} have been addressed using few-level configurations as model systems. The strong nonlinearities in these systems are also interesting from the perspective of the properties of heat transport. For instance, thermal rectification and amplification effects can be used for thermal control via all-thermal functionalities~\cite{li_negative_2006,wang_thermal_2007,li_colloquium_2012} such as diodes~\cite{segal_spin_2005,segal_single_2008,ojanen_selection_2009,
ruokola_thermal_2009,ruokola:2011,schaller_collective_2016,man_controlling_2016,ordonez-miranda_quantum_2017,barzanjeh_manipulating_2018,kargi_two_2019,bhandari_thermal_2021,
xu_heat_2021,poulsen_entanglement_2021,iorio_photonic_2021} and transistors~\cite{segal_nonlinear_2008,joulain:2016,zhang_coulomb_2018,guo_quantum_2018,majland_quantum_2020,tahir_minimal_2020}.

\begin{figure}[t]
\centering
\includegraphics[width=0.85\linewidth]{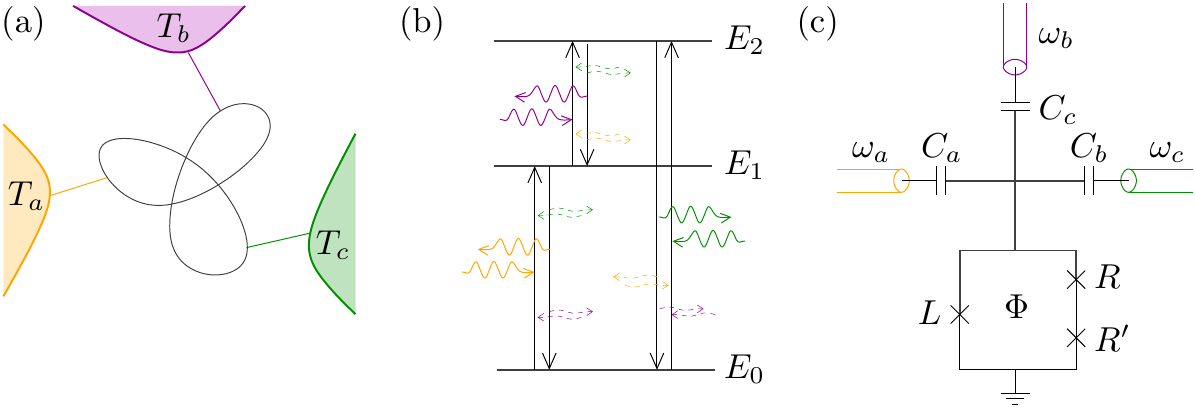}
\caption{(a) Scheme of a quantum system coupled to three reservoirs, each at a different temperature, $T_l$. (b) The three baths induce transitions between the three states $|0\rangle$, $|1\rangle$ and $|2\rangle$ with energies $E_i$. Each transition is predominantly induced by a single reservoir: bath $a$ is coupled to transition $|0\rangle\leftrightarrow|1\rangle$, $b$ to $|1\rangle\leftrightarrow|2\rangle$, and $c$ to $|0\rangle\leftrightarrow|2\rangle$. Residual (off-resonant) transitions induced by the other baths are marked with dashed arrows. 
(c) Proposed implementation of the system based on a cQED arquitecture. A Cooper pair box connected to a superconducting loop containing three Josephson junctions ($L$, $R$ and $R^\prime$) and threaded by a magnetic flux $\Phi$ defines the three-level system. It is capacitively coupled to three resonators at frequencies $\omega_l$ which mediate the coupling the thermal baths and filter the qutrit transitions.}\label{fig:scheme}
\end{figure}

Nonequilibrium operations require that the system is selectively coupled to several thermal baths at different temperatures, as sketched in Fig.~\ref{fig:scheme}(a),(b). In the ideal case, each reservoir drives a different transition in the system. This is possible e.g., in spatially-resolved configurations with the system being composed of smaller interacting parts, each of them connected to different reservoirs. Alternatively, single quantum systems with multiple levels can be considered where the system-bath couplings need to be filtered.
Both cases impose serious experimental limitations in situations requiring multiple baths, though some realizations of discrete-state autonomous heat engines have been achieved in three-terminal electronic setups~\cite{entin:2010,hotspots,thierschmann:2015,koski:2015,dorsch:2020} and trapped ions~\cite{maslennikov:2019}. 

Remarkably, considerable advances have been made in this direction in the field of circuit quantum electrodynamic (cQED) devices~\cite{vool_introduction_2017,blais_circuit_2021}. Nonlinearities in superconducting loops with Josephson junctions allow for the definition of tunable qubit or qutrit configurations with well resolved frequencies~\cite{krantz_quantum_2019}. 
Qutrits have been used in experimental realizations of quantum optics~\cite{baur_measurement_2009,sillanpaa_autler_2009,bianchetti_control_2010,kelly_direct_2010,abdumalikov_electromagnetically_2010,honiglDecrinis_mixing_2018}, quantum simulators~\cite{tan_topological_2018,vepsalainen_simulating_2020}, quantum information protocols~\cite{fedorov_implementation_2012,abdumalikov_experimental_2013,yurtalan_implementation_2020,morvan_qutrit_2021,cervera-Lierta_experimental_2021}, and quantum foundations~\cite{jerger_contextuality_2016} in superconducting circuits. 
The thermal baths can then be defined by external resistors whose coupling to the system is mediated via coplanar waveguide resonators~\cite{palacios-Laloy_tunable_2008,goppl_coplanar_2008}. This introduces the advantage that the baths can be spatially separated, even over macroscopic distances~\cite{partanen_quantum_2016}, allowing for good control of the temperature differences. Furthermore, the resonators act as natural filters for the bath-induced transitions in the system. These setups also allow for the measurement of heat currents via precise thermometry~\cite{giazotto:2006,pekola_colloquium_2021}. Quantum refrigerators~\cite{chen_quantum_2012,hofer_autonomous_2016}, heat engines~\cite{hofer_quantum_2016}, masers~\cite{thomas_thermally_2020} or thermal switches~\cite{karimi_coupled_2017} have been proposed based on these kinds of architectures, with recent implementation of qubit based heat valves~\cite{ronzani_tunable_2018} and thermal diodes~\cite{senior_heat_2020}. Remarkably, experimental realizations of three-terminal heat transport in a superconductor qutrit have recently been performed, achieving the measument of phased dependent heat current modulations of a few aW~\cite{gubaydullin_photonic_2021}.

Typically in the literature, perfect filtering is assumed such that each transition in the system is induced by the coupling to only one of the baths. Here we address the effect of imperfect filtering on the thermal conduction properties of the system~\cite{sanchez_single_2017,strasberg:2018,kargi_two_2019,tahir_minimal_2020}. This is indeed the case in most experiments, due to the finite width of the resonator spectral functions. Furthermore this is, up to some extent, a desirable property for practical reasons in the lab: very narrow resonances are difficult to match to the system frequencies~\cite{thomas_thermally_2020}. We consider a simple configuration consisting on a qutrit coupled to three thermal baths via three finite quality-factor resonators, see Fig.~\ref{fig:scheme}(c).
This results in heat leakage due to off-resonant transitions which are detrimental for the optimal performance of the system as a heat engine (either a refrigerator or a heat pump).
However these in principle undesired transitions introduce unexpected effects in the propagation of heat through the device. The thermal rectification properties of the qutrit are robust to heat leakage, which can even give rise to states with a perfect diode effect. It also introduces a thermal circulation~\cite{hwang:2018,acciai_phase_2021} effect that is not operative with ideally filtered couplings. 

The remaining of the text is organized as follows. Section~\ref{sec:model} presents the theoretical framework, both for a simple model with ideal couplings and the cQED-based proposal. The heat currents are analysed in Sec.~\ref{sec:currents}, and the resulting thermal diode and circulator operations are presented in Sec.~\ref{sec:diode}, before discussing the conclusions in Sec.~\ref{sec:conclusions}.

\section{The qutrit model and its implementation}
\label{sec:model}

We consider a simple three-level system whose states are labelled as $|i\rangle$, with $i=1,2,3$, described by a hamiltonian 
\beq
\label{eq:Hs}
\hat H_s=\sum_iE_i|i\rangle\langle i|. 
\eeq
The level energies, $E_i$, define the intrinsic frequencies of the system, $\omega_{ij}=(E_i-E_j)/\hbar$. 
The system is coupled to three thermal baths $l=a,b,c$ (though we will alleviate this condition eventually, see Sec.~\ref{sec:2termrect}) at temperatures $T_l$, that induce incoherent transitions between the different states, as sketched in Fig.~\ref{fig:scheme}(b). 

In the absence of filters, each bath $l$ can in principle induce all transitions in which the system changes its state from $|i\rangle$ to $|j\rangle$. This happens with a rate $\Gamma_{ji}^l$ that will be specified later in the different cases. In the weak system-bath coupling regime, higher order transitions including those simultaneously mediated by two baths can be neglected. With these assumptions, we can define the total rates $\Gamma_{ji}=\sum_l\Gamma_{ji}^l$ which will determine the system evolution via a quantum master equation for the system reduced density matrix, $\hat\rho$~\cite{breuer:book,gernot}. We are interested in the steady state dynamics, given by stationary solution of the master equation. In this case, it is given by a simple system of rate equations:
\beq
\label{eq:mastereq}
\sum_j\left(\Gamma_{ij}p_j-\Gamma_{ji}p_i\right)=0,
\eeq
where $p_i=\langle{i}|\hat\rho|{i}\rangle$ is the population of state $|i\rangle$. Coherences (nondiagonal elements) play no role in our description.

Once we know the populations, we obtain the heat current out of bath $l$ from:
\beq
\label{eq:jl}
J_l=\sum_{i,j}\hbar\omega_{ij}\Gamma_{ji}^lp_i.
\eeq 
Energy conservation involves $J_a+J_b+J_c=0$.
Note we have not done any assumption so far on what bath induces which transition.

Let us for later convenience introduce the notation convention that $J_{l,(m...n)}$ denotes the heat current in terminal $l$ when terminals $m,...,n$ are at temperature $T_h=T+\Delta T$ and all other terminals are at temperature $T$.

\subsection{Perfectly filtered couplings}
\label{sec:ideal}

Let us discuss first the case where the system-bath couplings are perfectly filtered, such that each bath only exchanges photons of a particular frequency, $\omega_l$, with the system. If these frequencies match those of the system, each bath will couple to a single transition only. This is the case for instance if $\omega_a=\omega_{10}$, $\omega_{b}=\omega_{21}$, and $\omega_c=\omega_a+\omega_b$. Then the only finite excitation rates are $\Gamma_{10}^a=\kappa_a\nB(\omega_a,T_a)$, $\Gamma_{21}^b=\kappa_b\nB(\omega_b,T_b)$, and $\Gamma_{20}^c=\kappa_c\nB(\omega_c,T_c)$, with the coupling parameters $\kappa_l$ and the Bose-Einstein distribution function
\beq
\nB(\omega,T)=\left[\rmee^{\hbar\omega/\kBT}-1\right]^{-1}.
\eeq
The reversed relaxation transitions are related to those by local detailed balance: 
\beq
\label{eq:detailbal}
\Gamma_{ij}^l=\Gamma_{ji}^l\rmee^{\hbar\omega_{ji}/\kBT_l}.
\eeq

With these, we solve the master equation \eqref{eq:mastereq}. Assuming for simplicity symmetric couplings $\kappa\equiv\kappa_1=\kappa_2=\kappa_3$, we obtain simple expressions for the heat currents:
\beq
\label{eq:idealJ}
J_l^{\rm f}=s_l\omega_l{\cal A}\left(\frac{\hbar\omega_a}{\kBT_a},\frac{\hbar\omega_b}{\kBT_b},\frac{\hbar\omega_c}{\kBT_c}\right),
\eeq
with $s_a=s_b=1$ and $s_c=-1$, and
\beq
\label{eq:calA}
{\cal A}(\theta_a,\theta_b,\theta_c)=\kappa\frac{\rmee^{\theta_a{+}\theta_b}-\rmee^{\theta_c}}{2+2\rmee^{\theta_b}+\rmee^{\theta_c}
-\rmee^{\theta_a{+}\theta_b}-2\rmee^{\theta_b{+}\theta_c}
-\rmee^{\theta_a{+}\theta_b+\theta_c}}.
\eeq
The superscript f in Eq.~\eqref{eq:idealJ} emphasizes that it holds only for perfect filtering. The denominator in Eq.~\eqref{eq:calA} is positive, so all the thermodynamic properties are given by the numerator. Note that all three currents are tightly coupled by the function ${\cal A}=\omega_a^{-1}J_a^{\rm f}=\omega_b^{-1}J_b^{\rm f}=-\omega_c^{-1}J_c^{\rm f}$. This is because, in order for heat to be transported through the system, the {\it basic} cycle 
\beq
\label{eq:basiccycle}
|0\rangle\stackrel{a}{\longleftrightarrow}|1\rangle\stackrel{b}{\longleftrightarrow}|2\rangle\stackrel{c}{\longleftrightarrow}|0\rangle
\eeq
needs to be completed, with each transition involving a different reservoir. This is the case also for asymmetric couplings.
This property has important consequences, as it involves that all currents vanish at a single point (other than equilibrium) given by:
\beq
\label{eq:stallT}
\frac{\hbar\omega_c}{\kBT_c}=\frac{\hbar\omega_a}{\kBT_a}+\frac{\hbar\omega_b}{\kBT_b}.
\eeq
From this expression we also learn that if only one of the baths is out of equilibrium, say with a temperature $T_H=T+\Delta T$, the system will necessarily exchange heat with all other baths. There is no way that all heat injected from the hot bath is absorbed by a single other one. Furthermore, we note that $J_a$ and $J_b$ have the same sign. This property enables the system to work as a refrigerator when one of these two terminals is hot~\cite{kosloff_quantum_2014}.

We can also check the cyclic relation:
\beq
\label{eq:jcycl}
J_{a,b}J_{b,c}J_{c,a}-J_{a,c}J_{c,b}J_{b,a}=0
\eeq
that will become relevant later.

\subsection{Circuit QED arquitecture}
\label{sec:cqed}

Let us from now on consider the system to be realized by a superconducting qutrit, following the proposal in Ref.~\cite{thomas_thermally_2020}, as represented in Fig.~\ref{fig:scheme}(c). The qutrit is defined by a superconducting island connected via the two arms of a loop to the superconducting ground. The loop contains three Josephson junctions, one in the left ($L$) and two in the right arm ($R$ and $R^ \prime$). The island charging energy is given by $E_C=(2e)^2/2C_\Sigma$ (in terms of the total capacitance $C_\Sigma=\sum_lC_l+C_L+C_R$, including those of the Josephson junctions)~\cite{clarke_superconducting_2008}. For simplicity, we assume all Josephson junctions to have the same Josephson energy $E_J=\hbar I_0/2e$, defined by their critical current, $I_0$. We will assume that $E_J\gg E_C$. 

The current flowing through the loop depends on the phase difference across every junction, $\varphi_L=\varphi$ and $\varphi_R=\varphi_{R^\prime}=\varphi^\prime/2$. They are related by the phase shift $\phi=(2e/\hbar)\Phi$ induced by the magnetic flux $\phi$: $\phi=\varphi-\varphi^\prime$. The current hence reads:
\beq
I(\varphi)=I_0\left(\sin\frac{\varphi-\phi}{2}+\sin\varphi\right).
\eeq
With this, we obtain the potential across the loop,
\beq
U(\varphi)=\frac{\hbar}{2e}\int{d\bar\varphi}I(\bar\varphi).
\eeq
Then, the hamiltonian of the system, after considering the charge contribution, is given by:
\beq
\hat H_s=-4E_C\partial_\varphi^2+E_J\left(3\cos\frac{\phi}{3}-\cos\varphi-2\cos\frac{\varphi-\phi}{2}\right),
\eeq
which has the nice property that the potential is zero at the minimum $\varphi=\phi/3$. Expanding around this condition, we get 
\beq
\hat H_s=-4E_C\partial_\varphi^2+\frac{1}{2}\tilde E_J\varphi^2-\frac{1}{8}E_J\sin\frac{\phi}{3}\varphi^3-\frac{1}{32}\tilde E_J\varphi^4+{\cal O}\left(\varphi^5\right),
\eeq
where $\tilde E_J=(3/2)E_J\cos(\phi/3)$. In the first two terms one recognizes a harmonic oscillator with frequency $\omega_0=\sqrt{8\tilde E_JE_C}/\hbar$, while the cubic and quartic terms are responsible for the system anharmonicity that allows for the definition of the qutrit. 

The system energies are obtained by perturbation theory on the anharmonic terms, resulting in:
\beq
E_n\approx\hbar\omega_0\left(n+\frac{1}{2}\right)-\frac{1}{16}E_C(6n^2+6n+3).
\eeq
The three lower states will form the qutrit, fixing the frequencies
$\omega_{10}=\omega_0-3E_C/4\hbar$, and $\omega_{21}=\omega_0-3E_C/2\hbar$. They can however be tuned externally via the magnetic flux (implicit in $\omega_0$). Note that the frequency $\omega_{32}$ sets an upper bound for the width of the resonator spectral functions, if one does not want to start populating higher excited states. With these energies we write the total hamiltonian, $\hat H=\hat H_s+\hat H_r+\hat H_{s-r}$, where $\hat H_s$ takes the same form as Eq.~\eqref{eq:Hs}, the resonators are described by
\beq
\label{eq:Hres}
\hat H_{r}=\sum_l\hbar\omega_l\left(\hat a_l^\dagger\hat a_l+\frac{1}{2}\right),
\eeq
and the system-resonator coupling is:
\beq
\label{eq:Hsysres}
\hat H_{s-r}=\sum_{ij,l}\hbar g_{ij}^l\left(\hat a_l^\dagger+\hat a_l\right)\hat X_{ij},
\eeq
with $\hat X_{ij}=|i\rangle\langle j|+|j\rangle\langle i|$. The coupling constants depend on the system wavefunctions~\cite{thomas_thermally_2020}, as well as on the resonator parameters (their capacitance, $C_l$, resistance, $R_l$, and inductance, $L_l$), 
which also fix the resonator frequency $\omega_l=1/\sqrt{L_lC_l}$ and the impedance $Z_0^l=\sqrt{L_l/C_l}$.
The resonator impedances introduce the fluctuating environment responsible for dissipation. In the weak-coupling limit, the transition rates are given by the Fermi golden rule. They are proportional to the voltage noise of the resonator:~\cite{ronzani_tunable_2018}
\beq
S^l(\omega)=2R_l\hbar\omega\left[1+Q_l^2\left(\frac{\omega}{\omega_l}-\frac{\omega_l}{\omega}\right)^2\right]^{-1}[1+\nB(\omega,T)],
\eeq
with the resonator quality factors $Q_l=Z_0^l/R_l$.
For our purposes here, it is enough to capture the dependence on $Q_l$ in the transition rates by introducing the dimensionless parameter $\lambda_{ij}^l\propto|g_{ij}^l|^2$, leading to
\beq
\label{eq:rates}
\Gamma_{ji}^l=\lambda_{ij}^l\frac{2\omega_{ji}}{Q_l}\left[1+Q_l^2\left(\frac{\omega_{ji}}{\omega_l}-\frac{\omega_l}{\omega_{ji}}\right)^2\right]^{-1}\nB(\omega_{ji},T_l).
\eeq
For more detailed discussions in terms of the experimental setup parameters, see Refs.~\cite{ronzani_tunable_2018,thomas_thermally_2020}. With these rates, we are ready to write the master equation \eqref{eq:mastereq} and obtain the dissipated heat currents using Eq.~\eqref{eq:jl}. Note that the rates in Eq.~\eqref{eq:rates} verify detailed balance as in Eq.~\eqref{eq:detailbal}.

\subsection{Conventions}
\label{sec:conv}

In the following we will consider a configuration in which the frequency of each resonator perfectly matches that of one of the system transitions, in particular $\omega_a=\omega_{10}$, $\omega_b=\omega_{21}$, and $\omega_c=\omega_{20}$. Hence these transitions will be predominantly induced by energy exchange with the corresponding bath. Due to the frequency dependence of the transition rates, however, the coupling is not restrictive: each bath is in principle able to induce all other system transitions with a finite rate, which increases for smaller resonator quality factors. We will differentiate them by assuming $\lambda_{10}^a=\lambda_{21}^b=\lambda_{20}^c=\lambda_{\rm res}$, and $\lambda_{ij}^l=\lambda_{\rm off}$, otherwise. Off-resonant transitions are represented with dashed arrows in Fig.~\ref{fig:scheme}(b). The ideal case discussed in Sec.~\ref{sec:ideal} is recovered for asymptotically high $Q_l$ or by making $\lambda_{\rm off}=0$. We will consider all reservoirs to have the same quality factors $Q_l=Q$ and, except when explicitly stated, $\lambda_{res}=\lambda_{\rm off}\equiv\lambda$ (deviations from this simplification affect the relative magnitude of the currents but do not change our main conclusions). We will furthermore consider a reference frequency $\omega_r/2\pi=\unit[1]{GHz}$ for frequencies, $\hbar\omega_r$ for energies, and $\lambda\hbar\omega_r^2$ for currents.

\section{Heat currents}
\label{sec:currents}

Transitions in the qutrit involve energy to be absorbed from, or emitted into one of the thermal baths. Having three baths that can in principle be at different temperatures makes the heat currents flow in a non trivial way (in the sense of not just going from hot to cold, as is the case with just two baths). 
For instance, consider the case that $a$ and $b$ are respectively the hottest and the coldest baths, i.e., $T_a>T_c\ge T_b$. They are predominantly coupled to transitions $|0\rangle\rightarrow|1\rangle$ and $|1\rangle\rightarrow|2\rangle$, so the population of $|1\rangle$ will be enhanced by thermal fluctuations. In this situation, and for moderate temperature differences $T_c-T_b$, transitions involving bath $b$ will most likely involve the absorption of an energy $\hbar\omega_{21}$ from it, resulting in cooling, $J_b>0$. For large differences between the cold and room reservoirs, $T_c-T_b$, the operation is reversed, so heat is injected into bath $a$. Hence the system works as a heat pump. Similar arguments when $T_b>T_c\ge T_a$ lead to the possibility of cooling reservoir $a$ or of pumping into $b$. 

\begin{figure}[t]
\centering
\includegraphics[width=0.85\linewidth]{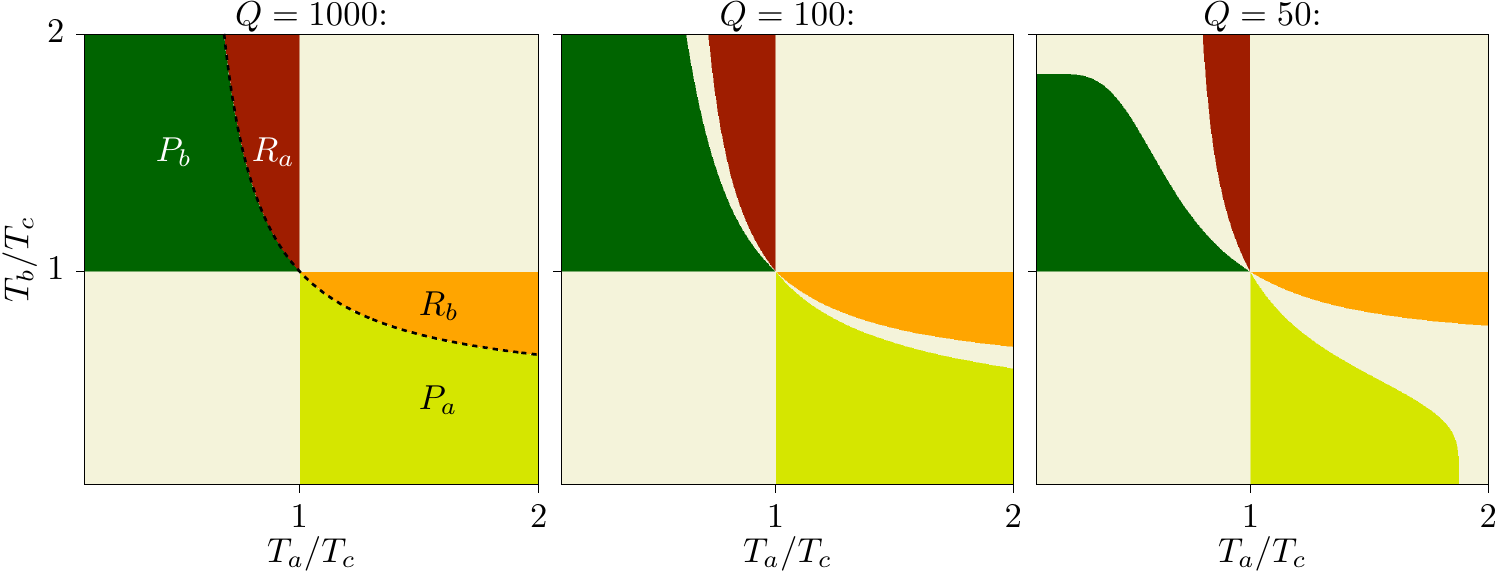}
\caption{Effect of the resonator quality factors on the thermodynamic operations of the qutrit: $R_l$ for the refrigeration of bath $l$, and $P_l$ for heat pumping into $l$. Here: $E_J=5\hbar\omega_r$, $E_C=0.5\hbar\omega_r$, $\omega_a=\omega_{10}$, $\omega_b=\omega_{21}$, and $\omega_c=\omega_{20}$. Lower quality factors ($Q\sim10$) would imply that higher energy levels start to become populated. The dashed line marks the condition for vanishing currents in the ideal case $g_l'=0$.}\label{fig:opermap}
\end{figure}

The regimes where these two operations occur in our system are shown in Fig.~\ref{fig:opermap} as a function of the temperatures of baths $a$ and $b$, with $T_c$ fixed. In the high-$Q$ regime, the vanishing current condition set by Eq.~\eqref{eq:stallT}, and marked by a black-dashed line in Fig.~\ref{fig:opermap}(a), separates regions where one of reservoirs $a$ or $b$ is cooled (marked as $R_a$ and $R_b$, respectively) from those where heat is pumped in the other one (marked as $P_b$ and $P_a$). In the absence of a work source, no hybrid operation (e.g., simultaneously cooling and pumping) is possible~\cite{manzano_hybrid_2020}.

\begin{figure}[t]
\centering
\includegraphics[width=0.85\linewidth]{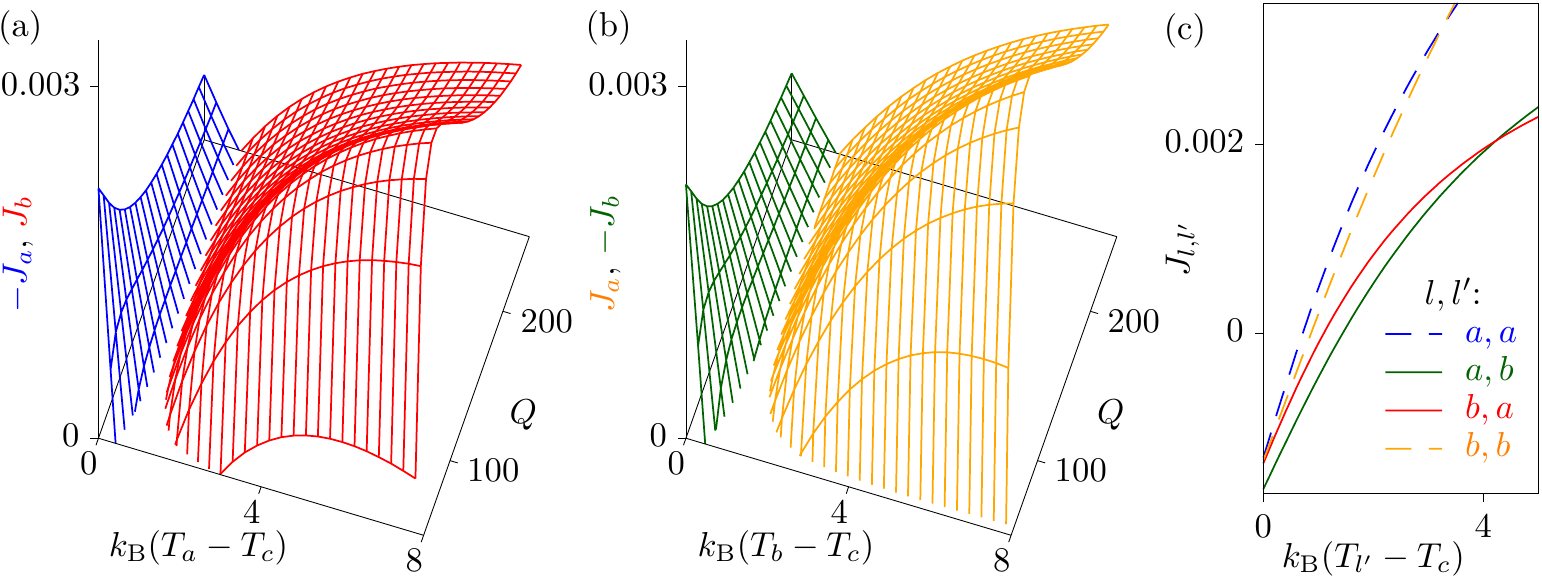}
\caption{Dependence of the cooling power and heat pumping on the resonator quality factors, as functions of the hot bath temperature in different configurations: (a) $T_a\ge T_c>T_b$, with fixed $\kBT_b=1.5$, so $J_a<0$ involves heat pumping in bath $a$, and $J_b>0$, cooling of bath $b$; and (b) $T_b\ge T_c>T_a$, with fixed $\kBT_a=1.5$, so $b$ is pumped and $a$ is cooled down. In both cases, $\kBT_c=2$. Currents are plotted only in the regions of useful operation. Panel (c) shows the different heat currents for the case with $Q=100$. Same parameters as in Fig.~\ref{fig:opermap}. }
\label{fig:operTHQ}
\end{figure}

As the quality factors become smaller, off-resonant transitions start to contribute, modifying the perfectly cyclic transitions of the ideally filtered case, cf. Eq.~\eqref{eq:basiccycle}. The residual heat leakage breaks the tight-coupling relation of $J_a$, $J_b$ and $J_c$ expressed by Eqs.~\eqref{eq:idealJ} and \eqref{eq:calA}. As a consequence, they do not vanish at the same condition, such that the cooling and pumping regions become separated and shrank, see Figs.~\ref{fig:opermap}(b) and (c). The operation efficiencies are reduced by the leakage currents, however imperfect filtering can be beneficial for increasing the pumped currents, see $-J_a$ in Fig.~\ref{fig:operTHQ}(a) and $-J_b$ in \ref{fig:operTHQ}(b). Also for the refrigerator, the cooling power has an optimal cooling factor for intermediate values of $Q$, see $J_b$ in Fig.~\ref{fig:operTHQ}(a) and $J_a$ in \ref{fig:operTHQ}(b). 
Note that this efficiency-power trade-off in the presence of leakage currents due to broad resonances is common to other types of heat engines~\cite{humphrey:02,nakpathomkun:2010,jordan:2013}.
Remarkably, in the region that separates cooling and pumping, $J_a$ and $J_b$ have opposite signs, as shown in Fig.~\ref{fig:operTHQ}(c), a property that will become useful in the later discussion.


\section{Rectification}
\label{sec:diode}

Apart from the just discussed thermodynamic operations, most obviously linked to the second law (i.e., cooling and pumping), the properties of heat transport through multiterminal devices are of interest from the point of view of heat control in networks. For instance, thermal analogues of electrical diodes, transistors or circulators can be defined. In the following, we explore how the heat currents flow in each bath depending on which of them is acting as a heat source. In particular, we focus on situations for which the system acts as a thermal rectifier, both in three- and two-terminal configurations, and as a circulator. 

In a thermal rectifier, the (forward) current $J_{l,l^\prime}$ absorbed by reservoir $l$ when another bath, $l^\prime$, is hot, with $T_{l^\prime}=T+\Delta T\equiv T_H$ and $T_l=T$, is different from the opposite (backward) current $J_{l^\prime,l}$ absorbed by $l^\prime$ when $T_l=T_H$ and $T_{l^\prime}=T$. Then, heat currents are not reciprocal in $l$ and $l'$,  $J_{l,l^\prime}\neq J_{l^\prime,l}$. In two-bath systems, this effect requires the presence of broken inversion symmetry (via e.g., asymmetric couplings) and nonlinearities~\cite{benenti_from_2016}. The effect can also be defined in multiterminal systems if heat dissipated in the other reservoirs (which are neither $l$ nor $l'$) is not important to the problem. We will call these the {\it passive} reservoirs. 
Indeed, in some cases, additional degrees of freedom of the system can act as a passive third reservoir, making the multiterminal description appropriate. This occurs for instance in electronic devices, where inelastic scattering due to energy exchange with lattice phonons ~\cite{martinez-Perez_rectification_2015,jiang:2015,donald} or fluctuations of the electromagnetic environment~\cite{rossello:2017} are sometimes difficult to prevent. Note this is different from electrical diodes, which are purely two-terminal devices. 

\subsection{Three-reservoir rectification}
\label{sec:3termrect}

Unlike in two-terminal configurations, a three-terminal rectification effect can be found in the linear regime~\cite{rossello:2017,chiraldiode}.
The reason for this is that the third reservoir acts as a heat sink that breaks the conservation of currents in the involved terminals $J_l+J_{l^\prime}\neq0$. In our system, this third bath (the one that is neither $l$ nor $l^\prime$) is treated on equal footing as the other two, and can be independently controlled and measured. For simplicity, in this subsection we consider that it remains at temperature $T$ in the forward and backward states. We will allow it to have a different temperature in Sec.~\ref{sec:2grad}.

We quantify this effect by introducing the rectification coefficient as:
\beq
\label{eq:3tR}
{\cal R}_{ll^\prime}=-\frac{J_{l,l^\prime}-J_{l^\prime,l}}{|J_{l,l^\prime}|+|J_{l^\prime,l}|}.
\eeq
If the forward and backward currents are similar, ${\cal R}_{ll^\prime}\approx0$. If one of the currents is orders of magnitude larger than the other one, the rectification is large, ${\cal R}_{ll^\prime}\approx\pm1$ and the system behaves as a thermal diode for baths $l$ and $l'$. The minus sign in the definition is because heat flowing into the reservoirs is defined as negative.  

\begin{figure}[t]
\centering
\includegraphics[width=0.85\linewidth]{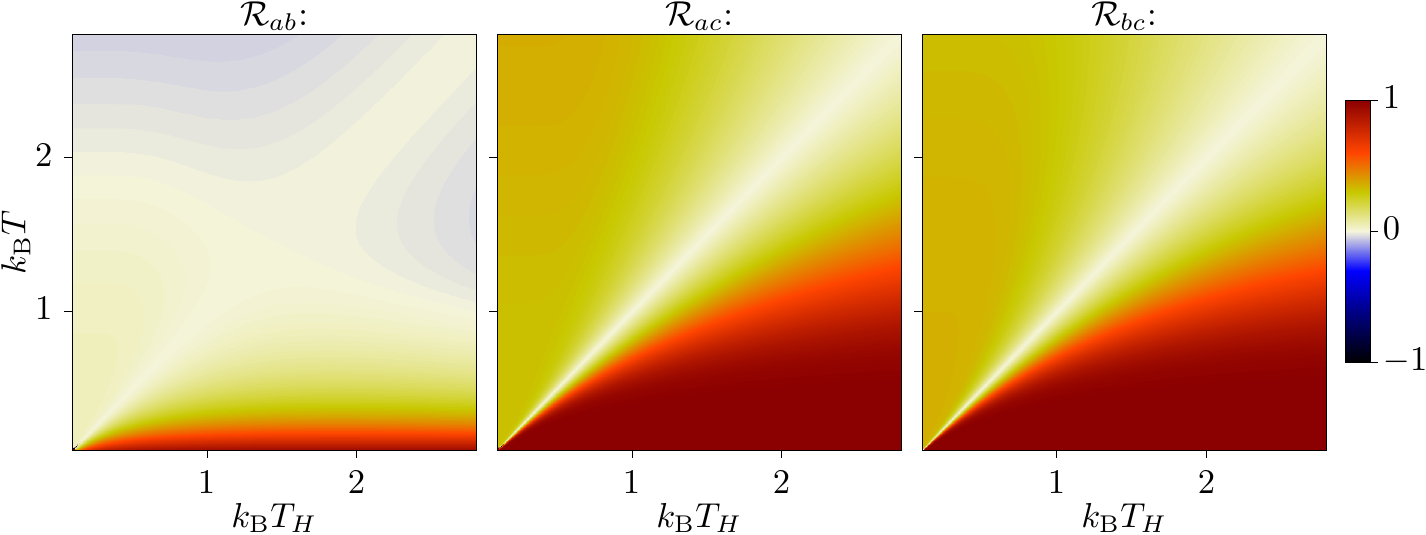}
\caption{Three-terminal rectification, in the perfect filtering case with $\lambda_{\rm off}=0$. The coefficients ${\cal R}_{ab}$, ${\cal R}_{ac}$ and ${\cal R}_{bc}$ are plotted as functions of the temperature $T$ and the tuned temperature $T_H$, for otherwise the same parameters as in Fig.~\ref{fig:opermap}.}
\label{fig:3trectg0}
\end{figure}

We calculate the three rectification coefficients, ${\cal R}_{ab}$, ${\cal R}_{ac}$ and ${\cal R}_{bc}$, for the perfect filtering case with $\lambda_{\rm off}=0$, and plot them in Fig.~\ref{fig:3trectg0}. They are plotted as functions of the base temperature $T$ and of the source temperature, $T_H$, emphasizing that the rectification effect also occurs when $T_H$ is actually colder than $T$ (so $\Delta T<0$). However, in all three configurations, the rectification is smaller for colder reservoirs, as compared to the positive $\Delta T$ case. In this case, where we explicitly considered all couplings to be the same, the asymmetry that generates the rectification is only due to the spectral properties of the system. The rectification is strongest when involving bath $c$, which is coupled to the largest system frequency, $\omega_{20}$, i.e., for $l=a,b$ and $l^\prime=c$, finding ${\cal R}_{ac},{\cal R}_{bc}\approx1$ when $\Delta T\gg T$. 

To understand this effect, consider for instance the rectification between baths $a$ and $c$. The remaining (passive) bath $b$ (always at temperature $T_b=T$) couples to the transitions between states $|1\rangle$ and $|2\rangle$. In the forward configuration, where $T_c=T_H$ (with $\Delta T>0$) and $T_a=T$, the hot bath excites the system from the ground up to state $|2\rangle$. The basic cycle is then completed by relaxing sequentially to $|1\rangle$, and back to $|0\rangle$ by emitting energy into the cold reservoirs, $b$ and $a$. On the contrary, for the backward configuration, when $T_c=T$ and $T_a=T_H$, the cycle is reversed and requires the absorption of energy from the reservoir $b$, which is cold. The transition $|1\rangle\rightarrow|2\rangle$ is hence exponentially suppressed [cf. Eq.~\eqref{eq:detailbal}], and so is the current $J_c$. Our system then works as a thermal diode. In the case with $\Delta T<0$, the effect can be understood in similar terms, but the performance is reduced.

A finite rectification effect also appears between terminals $a$ and $b$, see Fig.~\ref{fig:3trectg0}(a). However, the rectification coefficient is small except for very low $T$, where it tends toward ${\cal R}_{ab}\rightarrow1$, a regime that is in general difficult to achieve experimentally. Also, the currents become very small, unless the temperature difference is furthermore large, $\Delta T\gg T$.

\begin{figure}[t]
\centering
\includegraphics[width=0.85\linewidth]{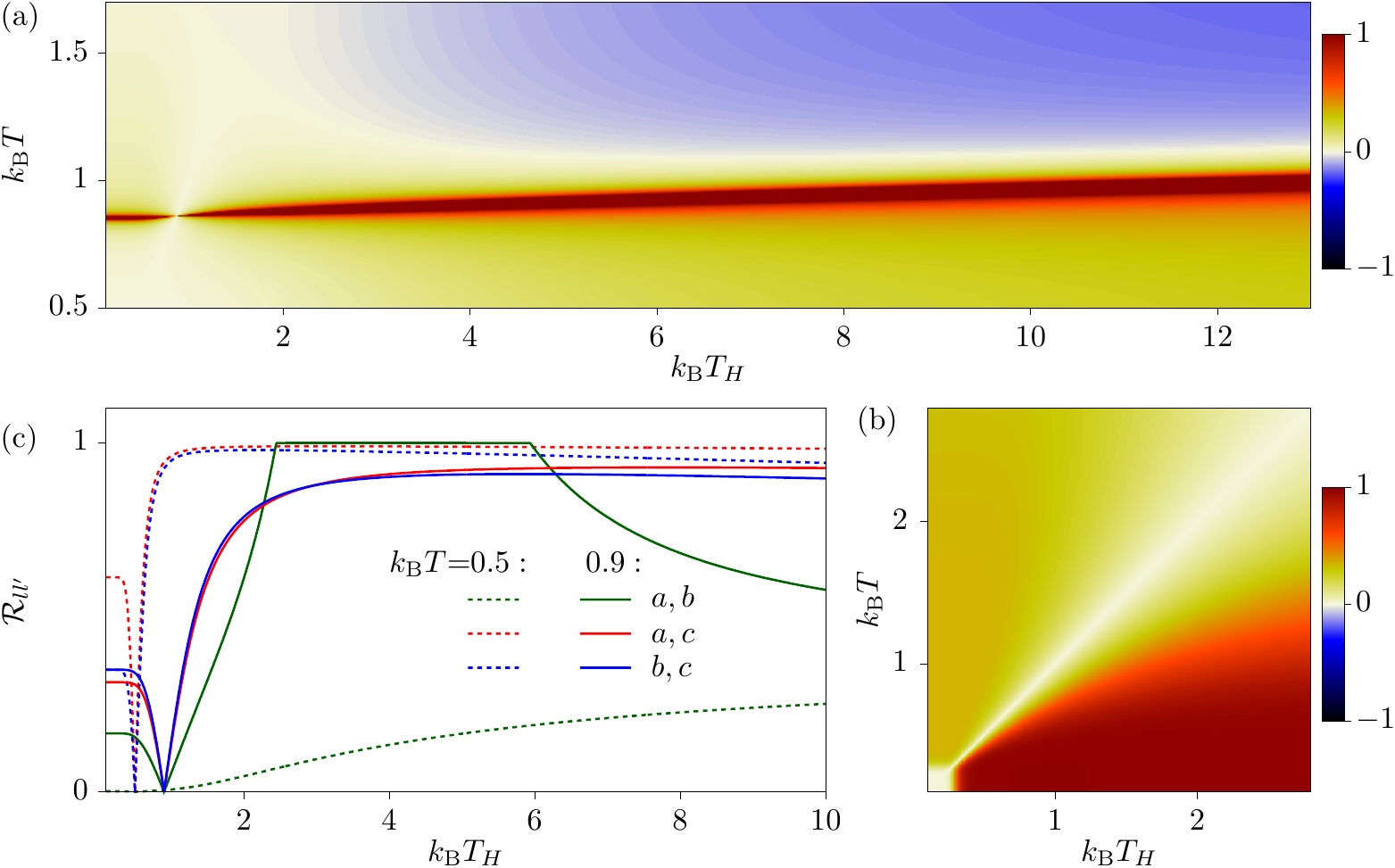}
\caption{Three-terminal rectification coefficients (a) ${\cal R}_{ab}$, and (b) ${\cal R}_{bc}$ as functions of $T$ and $T_H$.  (c) Cuts of all three rectification coefficients ${\cal R}_{ll^\prime}$ for different temperatures: $\kBT=0.5$ (solid lines) and $\kBT=0.9$ (dashed lines). Same parameters as in Fig.~\ref{fig:opermap}.}
\label{fig:3trect}
\end{figure}

The situation changes for finite quality factors, as shown in Fig.~\ref{fig:3trect}. The most dramatic case is indeed the rectification ${\cal R}_{ab}$, which we showed to be small when filtering is perfect. The reason is the change of the temperatures at which the different currents change sign. We showed in Fig.~\ref{fig:operTHQ}(c) that $J_{a,b}$ and $J_{b,a}$ vanish at different points. In between these two points, the two currents have opposite signs i.e., the configuration with $T_a=T_H>T$ is working as an absorption refrigerator for $b$ ($J_{b,a}>0$), while in the opposite one, reservoir $a$ absorbs part of the heat injected from a hot $b$ ($J_{a,b}<0$). Note that in both configurations heat is extracted from reservoir $b$, and injected into reservoir $a$ i.e., the heat flow between $a$ and $b$ has the same sign irrespective of which of them is the heat source. At the vanishing points, the system behaves as a perfect diode: one of the configurations work as a thermal insulator. In this region, we find ${\cal R}_{ab}=1$, as shown in Figs.~\ref{fig:3trect}(a) and (c). In the region with $T_H<T$, the same arguments hold for reversed currents, where the perfect diode effect occurs when pumping heat into reservoir $a$ (as we have $J_{a,b}<0$ when $T_b=T_H<T$).  
A somewhat related effect, also due to leakage currents close to tightly-coupled configurations, is predicted for heat flows in quantum dot systems~\cite{sanchez_single_2017}. It can also be found in two terminal systems under a time dependent modulation~\cite{carrega_engineering_2021}.

The other coefficients, ${\cal R}_{ac}$ and ${\cal R}_{bc}$ are weakly affected by the leakage currents, except for being strongly suppressed when both temperatures are very low, as shown in Fig.~\ref{fig:3trect}(b).
Changing the quality factor does not change the qualitative behaviour. The smaller $Q$, the larger is the range of temperatures $T_H$ for which the perfect diode operation with ${\cal R}_{ab}=1$ appears. It also occurs at larger temperatures $T$ for low quality factors. For very high $Q$, this behaviour shifts towards very low $T$, recovering the $\lambda_{\rm off}=0$ case shown in Fig.~\ref{fig:3trectg0}.

\subsection{Two temperature gradients}
\label{sec:2grad}

In the discussion above, we have fixed the temperature of the third reservoir (the passive sink) to be equal to the temperature of the coldest diode reservoir. However, this is not necessarily the case. Indeed, in experimental situations it might have a different temperature for being affected by a different environment or simply by adapting its temperature to the heat exchanged with the system~\cite{zhang_ballistic_2010,ming_ballistic_2010,donald}. Having all terminals at different temperatures has important consequences on the rectification properties. 

\begin{figure}[t]
\centering
\includegraphics[width=0.85\linewidth]{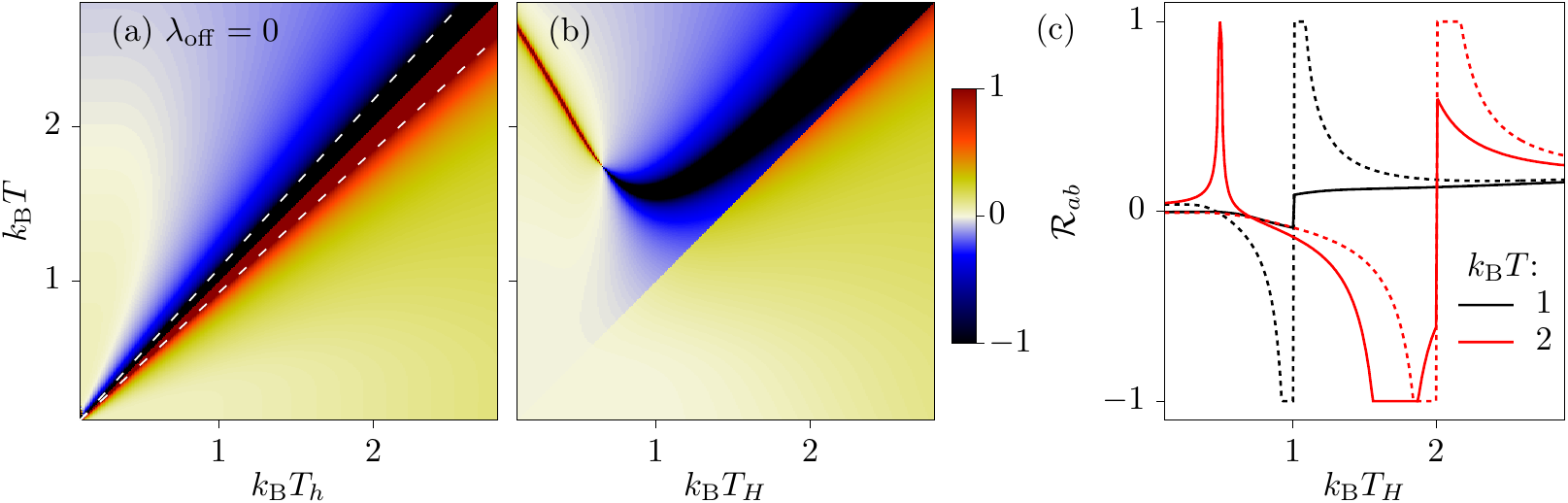}
\caption{Three-terminal rectification coefficient ${\cal R}_{ab}$ when the other terminal is at an intermediate temperature: $T_c=(T+T_H)/2$. (a) As $T$ and $T_H$ are varied, regions with ${\cal R}_{ab}=\pm1$ are found around $T_H=T$ for the perfectly filtered configuration. White dashed lines mark the conditions for $J_{a,b}=0$ and $J_{b,a}=0$. (b) For $Q=100$, the condition for vanishing currents is modified leading to a peak with ${\cal R}_{ab}=1$. Panel (c) shows cuts of for different $T$ of the cases in panels (a) in dashed lines and (b) in full lines. All other parameters are as in Fig.~\ref{fig:opermap}.}\label{fig:rectabTint}
\end{figure}

Let us consider for simplicity that the temperature of the passive reservoir is the mean of the other two: $T_c=T+\Delta T/2$. We focus on the case ${\cal R}_{ab}$. The results are plotted in Fig.~\ref{fig:rectabTint}, which shows that the rectification coefficient saturates to ${\cal R}_{ab}=\pm1$ also in the perfectly filtered configuration with $\lambda_{\rm off}=0$, see Fig.~\ref{fig:rectabTint}(a). This occurs in the region around the condition $T=T_H$ where all currents vanish. This region is bounded by (i) $T=T_H\omega_{10}/\omega_{21}$ and (ii) $T=T_H\omega_{21}/\omega_{10}$, where $J_{a,b}=0$ and $J_{b,a}=0$, respectively.  These bounds, established by Eq.~\eqref{eq:stallT}, are marked by white dashed lines in Fig.~\ref{fig:rectabTint}(a). In the region bounded by (i), the system pumps heat into reservoir $a$, while it pumps into reservoir $b$ in the region bounded by (ii).  

The lifting of tight-coupling for resonators with finite $Q$ changes the behaviour in a non-trivial way, as shown in Fig.~\ref{fig:rectabTint}(b). It also limits the temperatures at which pumping (which we just showed to be related to perfect rectification in this case) occurs, see also Fig.~\ref{fig:opermap}. There are lower bounds for the temperature at which we find $P_a$ and $P_b$ operations. The regions with ${\cal R}_{ab}=1$ and ${\cal R}_{ab}=-1$ are now split and separated from the equilibrium condition $T=T_H$, see Fig.~\ref{fig:rectabTint}(c). The transition from one region to the other takes place when $J_{a,b}=J_{b,a}=0$. Note that, in this particular case with a relatively low quality factor ($Q=100$), the optimal rectification is found in regions where $T_H<T$.

\subsection{Two-reservoir rectification}
\label{sec:2termrect}

With our model, we can also mimic a two-bath rectification effect if the system is modified such that two resonators share the same reservoir where to dissipate energy. The current into the bath connected to resonators $l$ and $l'$ is obtained from the expressions in Sec.~\ref{sec:model} by simply adding them, $J_l+J_{l^\prime}$ and taking care of having them at the same temperature, $T_l=T_{l^\prime}$. Under these conditions, the rectification coefficient is then:
\beq
\label{eq:2tR}
{\cal R}_{(ll^\prime)m}=-\frac{J_{l,m}+J_{l^\prime,m}-J_{m,(ll^\prime)}}{|J_{l,m}+J_{l^\prime,m}|+|J_{m,(ll^\prime)}|},
\eeq
where the indices between brackets indicate the resonators that are connected to the same bath. In this case, one of the baths couples to the system via two channels (resonators $l$ and $l'$), while the other one only via a single channel (resonator $m$). This is sufficient to introduce the asymmetry required for the rectification effect. Related configurations have been proposed that exploit selection rules~\cite{ojanen_selection_2009} or charge states in coupled quantum dots~\cite{alex}.

\begin{figure}[t]
\centering
\includegraphics[width=0.85\linewidth]{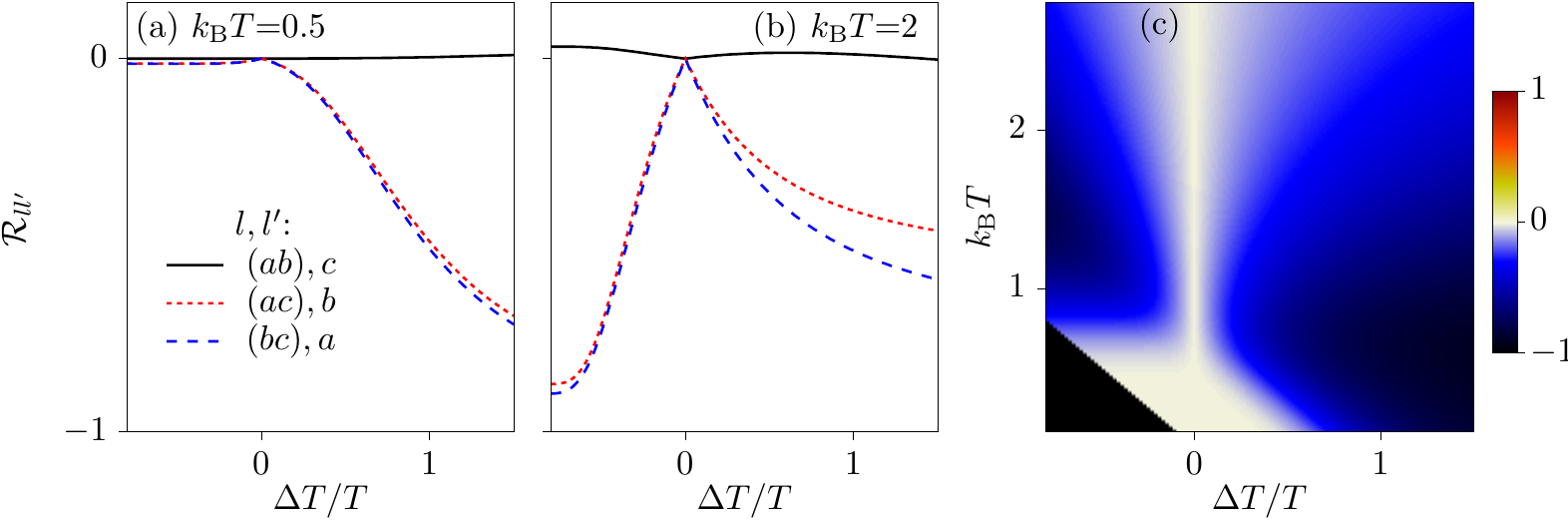}
\caption{Two-terminal rectification coefficients ${\cal R}_{(ll^\prime)m}$ as functions of (a), (b) $\Delta T=T_H-T$ (for fixed $\kBT=0.5$ and 2, respectively). (c) ${\cal R}_{(bc),a}$, resulting of connecting resonators $b$ and $c$ to the same thermal bath. The black region in the bottom-left corner corresponds to negative temperatures. All other parameters are as in Fig.~\ref{fig:opermap}.}\label{fig:2trect}
\end{figure}

Three different configurations are possible: ${\cal R}_{(ab),c}$, ${\cal R}_{(ac),b}$, and ${\cal R}_{(bc),a}$. Out of these, the fist one clearly results in the smallest rectification coefficient, as the system asymmetry is only due to the fact that the transitions between $|0\rangle$ and $|2\rangle$ occur in a single step by exchanging an energy $\hbar\omega_{20}$ with bath $c$, while it needs two subsequent transitions when due to energy absorbed from resonators $a$ and $b$, $\hbar\omega_{10}$ and $\hbar\omega_{21}$, respectively. In a basic cycle the two baths exchange the same amount of energy with the system. Deviations from ${\cal R}_{(ab),c}=0$ are then dominated by nonlinearities and slightly modulated by imperfect filtering, so the effect remains small, cf- Fig.~\ref{fig:2trect}(a) and (b). 

However, if the two-channel bath contains the highest frequency resonator, $c$, the asymmetry is maximized. After each basic cycle (with only resonant transitions), the two baths exchange the same amount of energy, however now the two-channel bath requires an absorption and an emission process. This leads to high values of ${\cal R}_{(ac),b}$ and ${\cal R}_{(bc),a}$, as shown for ${\cal R}_{(bc),a}$ in Fig.~\ref{fig:2trect}(c). Again, they approach $\pm1$ when $|\Delta T|$ is large compared to $T$. 

\subsection{Circulator}
\label{sec:circul}

A purely three-terminal device is the circulator, which makes current injected from one bath flow toward its nearest neighbours preferably in one direction, either clock- or counterclockwise. Despite their importance as electronic devices, few proposals for thermal circulators have been made so far~\cite{chiraldiode,hwang:2018,acciai_phase_2021}, mostly using heat carried by chiral states in electronic conduction. Here we propose an all-thermal circulator with no particle exchange. The circulation coefficient is defined as:
\beq
\label{eq:circcoeff}
{\cal C}=\frac{|J_\circlearrowright|-|J_\circlearrowleft|}{|J_\circlearrowright+J_\circlearrowleft|},
\eeq
with $J_\circlearrowright=J_{a,b}J_{b,c}J_{c,a}$ and $J_\circlearrowleft=J_{a,c}J_{c,b}J_{b,a}$. It is is bounded by $-1\leq{\cal C}\leq1$, when one of $J_\circlearrowright$ and $J_\circlearrowleft$ vanishes. Note that, different from the diode, the definition of the circulator is not based on the current of a single configuration, but rather on combinations of currents with all baths being subsequently hot.
In this case, the ideal filtering limit ($\lambda_l=0$) gives no circulation, ${\cal C}=0$, due to the cyclic property of Eq.~\eqref{eq:jcycl}. Any finite-circulation effect is hence due to the finite quality factors of the resonators. 

\begin{figure}[t]
\centering
\includegraphics[width=0.85\linewidth]{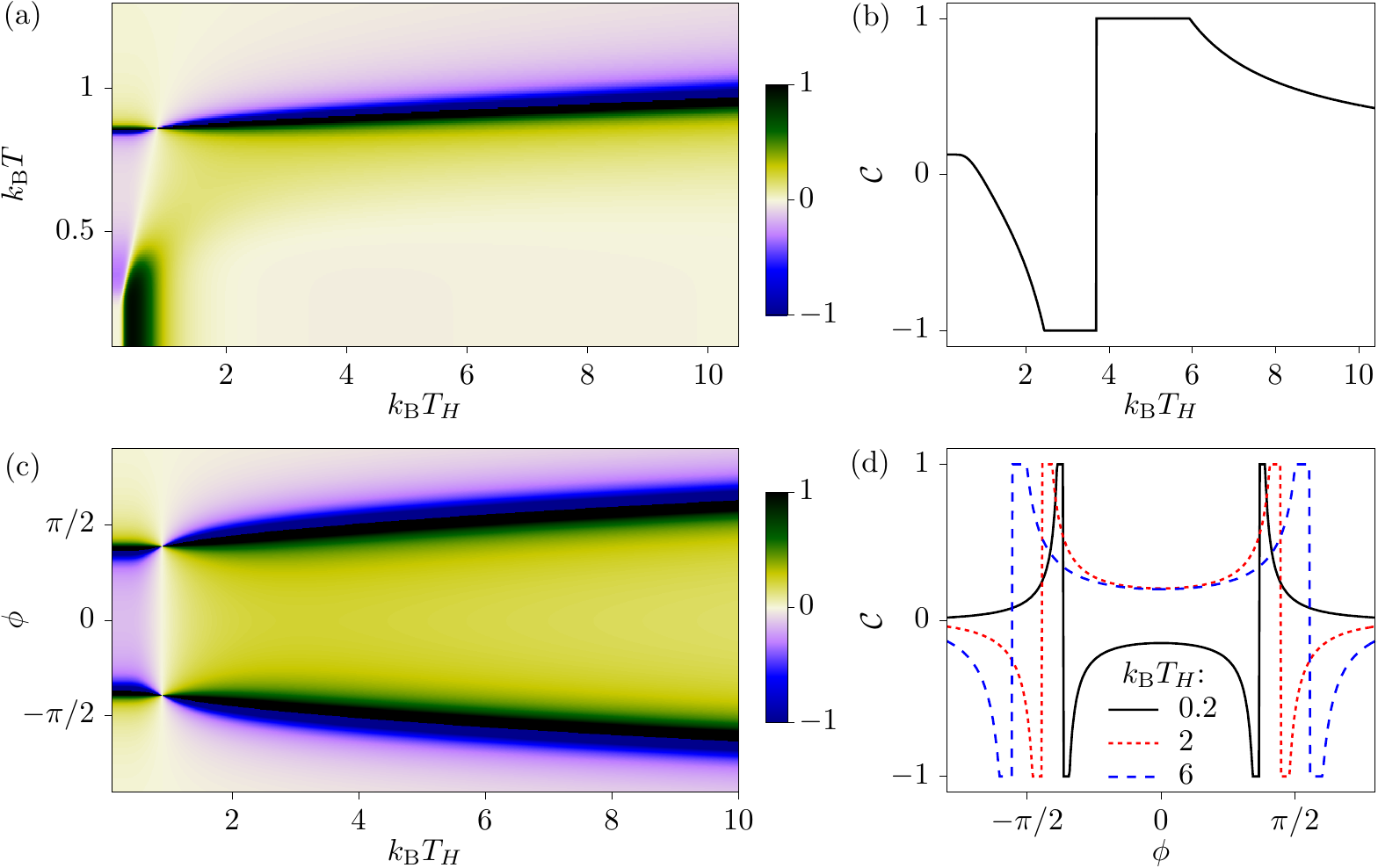}
\caption{(a) Thermal circulation coefficient, ${\cal C}$, as a function of $\kBT_H$ and  $\kBT$, and (b) a cut for fixed $\kBT=0.9$, both for $\phi=\pi/2$ and $Q_l=100$. (c) Dependence with the magnetic flux, fixing $\kBT=0.9$, and (d) for different $\kBT_H$ in the same configuration as (c). Parameters are as in Fig.~\ref{fig:opermap}. }\label{fig:circul}
\end{figure}

As for the rectifier, the desired properties are related to the occurrence of heat reversed current operations (cooling and pumping), combined with the leakage currents. In the region between the zeros of $J_{a,b}$ and $J_{b,a}$, cf. Fig.~\ref{fig:operTHQ}(c), the two currents have opposite contribution, resulting in the current circulating always in the same direction. This is indeed what we observe in Figs.~\ref{fig:circul}(a) and (b), where perfect circulation with ${\cal C}=\pm1$ is obtained in the mentioned region. In the borders of this region, $J_{a,b}$ and $J_{b,a}$ change sign, making either $J_\circlearrowright$ or $J_\circlearrowleft$ vanish. 

Interestingly taking advantage of the Josephson effect, both the sign and magnitude of the effect can be controlled by tuning the system frequencies with the flux $\phi$, as shown in Figs.~\ref{fig:circul}(c) and (d). This way one can switch between clockwise (${\cal C}=-1$) and counterclockwise (${\cal C}=1$) circulation states with a magnetic field.

\begin{figure}[t]
\centering
\includegraphics[width=0.85\linewidth]{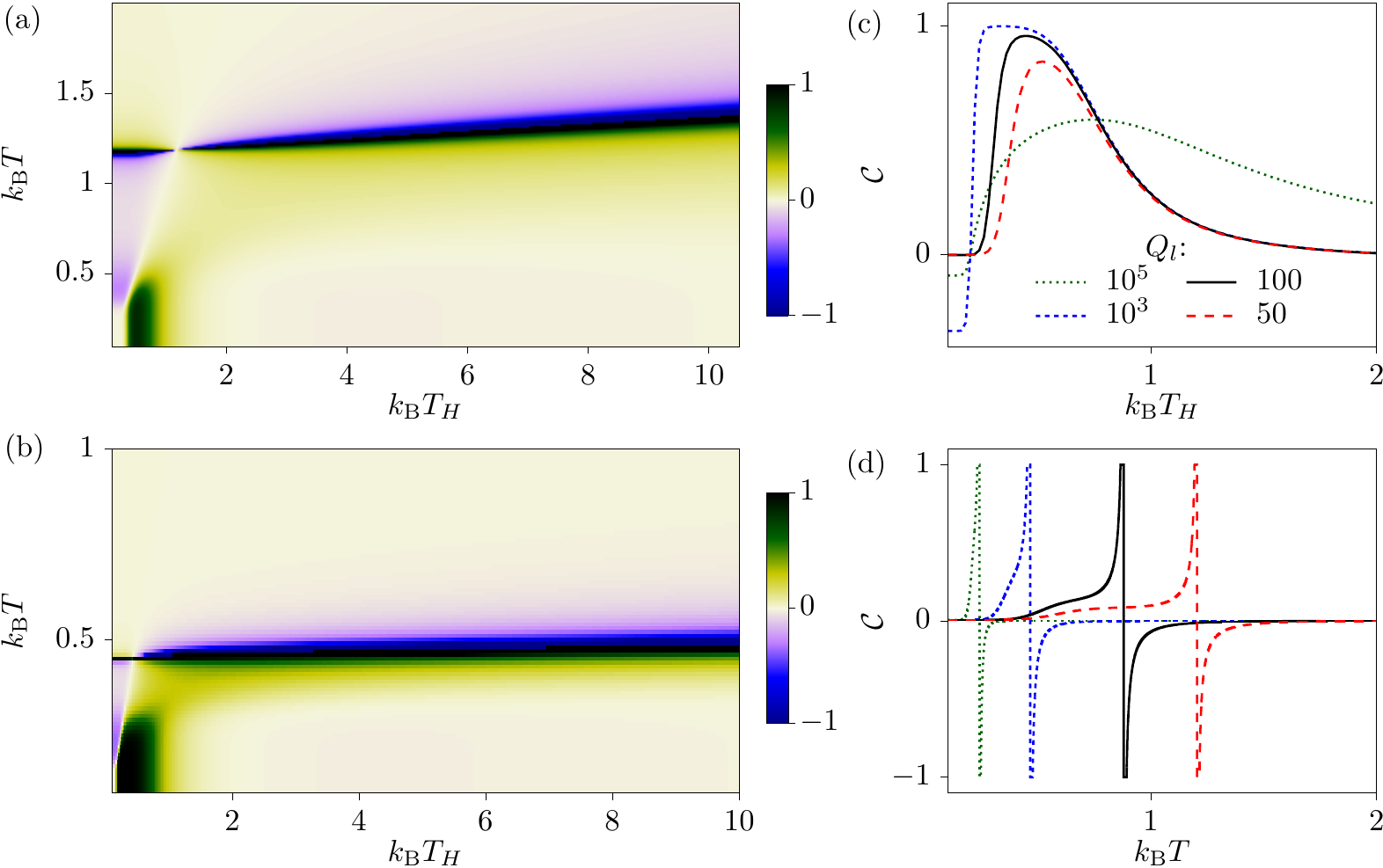}
\caption{Effect of the resonator quality factor on the thermal circulation coefficient. ${\cal C}$, as a function of $\kBT_H$ and  $\kBT$ for (a) $Q_l=50$, and (b) $Q_l=1000$. (c) Cuts at $\kBT=0.2$ for different quality factors. (d) Circulator coeffficient as a function of $\kBT$, for fixed $\kBT_H=2$ and the same $Q_l$ factors considered in (c). Parameters are as in Fig.~\ref{fig:circul}. }\label{fig:circulQ}
\end{figure}

We furthermore observe a region of relatively high ${\cal C}$ for low temperatures, where unfortunately currents are strongly suppressed. This feature increases with the quality factor, as shown in Fig.~\ref{fig:circulQ}(a) to (c), getting close to ${\cal C}=1$. However, for very high quality factors all the regions with perfect circulation behaviour move towards very low temperatures, see Fig.~\ref{fig:circulQ}(d). This can also be appreciated by comparing the position of the (almost) horizontal feature in Fig.~\ref{fig:circul}(a), for $Q=100$, with those in Figs.~\ref{fig:circulQ}(a) and (b), with $Q=50$ and $Q=1000$, respectively.

\section{Conclusions}
\label{sec:conclusions}

To summarize, we have explored the thermal properties of a qutrit coupled to three baths. For this, we propose a model based on recent experimental implementations of superconducting circuits, where the system-bath couplings are mediated by resonators. In this sense, our model goes beyond perfectly filtered transitions by allowing the resonators to have a finite quality factor that makes them influence transitions in the system off-resonantly.

We have shown that these in principle undesired transitions affect the heat transport properties. While they are expected to be detrimental for the efficiency of thermodynamic operations such as cooling and heat pumping, they can enhance their power. Most interestingly, they lift the condition for which all currents vanish in perfectly filtered models, resulting in situations in which the heat flow between two reservoirs has the same direction irrespective of which of them is hot. This effect is enabled by the current reversal effects (cooling, pumping) induced by the third reservoir. The qutrit then acts as a perfect thermal diode. 

We have explored different rectification effects, identifying different behaviours when allowing for the three baths to maintain one or more temperature differences (depending on whether the passive bath is at $T$ or at a different temperature). Two-reservoir configurations are also possible with enhanced rectification by introducing asymmetries in the number of system-bath coupling channels.

Finally, we have explored the thermal circulation properties of the system, finding that the same conditions that lead to the thermal rectification can be used to define an optimal and phase-tunable heat circulator.

These effects show that one can take advantage of experimental limitations to find useful thermal operations. Therefore, our results are not restricted to cQED setups but can be implemented in other few-level systems with non-perfectly filtered couplings to thermal baths, including optical cavities~\cite{gelbwaserKlimovsky_work_2015,kargi_two_2019,tahir_minimal_2020} or quantum dots~\cite{sanchez_single_2017,strasberg:2018}.  

\ack 
We thank Azat Gubaydullin for a useful discussion (even online poster sessions can be useful). We furthermore acknowledge him and Jukka P. Pekola for sharing the results of Ref.~\cite{gubaydullin_photonic_2021} with us prior to publication. R.S. acknowledges funding from the Ram\'on y Cajal program RYC-2016-20778, and the Spanish Ministerio de Ciencia e Innovaci\'on via grant No. PID2019-110125GB-I00 and through the ``Mar\'{i}a de Maeztu'' Programme for Units of Excellence in R{\&}D CEX2018-000805-M.


\section*{References}
\bibliographystyle{iopart-num}
\bibliography{biblio}

\providecommand{\newblock}{}
\begin{thebibliography}{100}
\expandafter\ifx\csname url\endcsname\relax
  \def\url#1{{\tt #1}}\fi
\expandafter\ifx\csname urlprefix\endcsname\relax\def\urlprefix{URL }\fi
\providecommand{\eprint}[2][]{\url{#2}}

\bibitem{scovil_three_1959}
Scovil H~E~D and Schulz-DuBois E~O 1959 {\em Phys. Rev. Lett.\/} {\bf 2}(6)
  262--263 \urlprefix\url{https://link.aps.org/doi/10.1103/PhysRevLett.2.262}

\bibitem{geusic_quantum_1967}
Geusic J~E, Schulz-DuBios E~O and Scovil H~E~D 1967 {\em Phys. Rev.\/} {\bf
  156}(2) 343--351
  \urlprefix\url{https://link.aps.org/doi/10.1103/PhysRev.156.343}

\bibitem{kosloff:1984}
Kosloff R 1984 {\em J. Chem. Phys.\/} {\bf 80} 1625
  \urlprefix\url{http://dx.doi.org/10.1063/1.446862}

\bibitem{mu_oneatom_1992}
Mu Y and Savage C~M 1992 {\em Phys. Rev. A\/} {\bf 46}(9) 5944--5954
  \urlprefix\url{https://link.aps.org/doi/10.1103/PhysRevA.46.5944}

\bibitem{kosloff_quantum_2014}
Kosloff R and Levy A 2014 {\em Annual Review of Physical Chemistry\/} {\bf 65}
  365--393 pMID: 24689798
  \urlprefix\url{https://doi.org/10.1146/annurev-physchem-040513-103724}

\bibitem{mitchison:2020}
Mitchison M~T 2019 {\em Contemp. Phys.\/} {\bf 60} 164
  \urlprefix\url{https://doi.org/10.1080/00107514.2019.1631555}

\bibitem{thermo:book}
Binder F, Correa L~A, Gogolin C, Anders J and Adesso G (eds) 2019 {\em
  Thermodynamics in the Quantum Regime\/} (Springer)
  \urlprefix\url{https://dx.doi.org/10.1007/978-3-319-99046-0}

\bibitem{palao_quantum_2001}
Palao J~P, Kosloff R and Gordon J~M 2001 {\em Phys. Rev. E\/} {\bf 64}(5)
  056130 \urlprefix\url{https://link.aps.org/doi/10.1103/PhysRevE.64.056130}

\bibitem{linden_how_2010}
Linden N, Popescu S and Skrzypczyk P 2010 {\em Phys. Rev. Lett.\/} {\bf
  105}(13) 130401
  \urlprefix\url{https://link.aps.org/doi/10.1103/PhysRevLett.105.130401}

\bibitem{levy_quantum_2012}
Levy A and Kosloff R 2012 {\em Phys. Rev. Lett.\/} {\bf 108}(7) 070604
  \urlprefix\url{https://link.aps.org/doi/10.1103/PhysRevLett.108.070604}

\bibitem{levy:2012pre}
Levy A, Alicki R and Kosloff R 2012 {\em Phys. Rev. E\/} {\bf 85}(6) 061126
  \urlprefix\url{https://link.aps.org/doi/10.1103/PhysRevE.85.061126}

\bibitem{correa_performance_2013}
Correa L~A, Palao J~P, Adesso G and Alonso D 2013 {\em Phys. Rev. E\/} {\bf
  87}(4) 042131
  \urlprefix\url{https://link.aps.org/doi/10.1103/PhysRevE.87.042131}

\bibitem{brunner_entanglement_2014}
Brunner N, Huber M, Linden N, Popescu S, Silva R and Skrzypczyk P 2014 {\em
  Phys. Rev. E\/} {\bf 89}(3) 032115
  \urlprefix\url{https://link.aps.org/doi/10.1103/PhysRevE.89.032115}

\bibitem{mohanta_universal_2021}
Mohanta S, Saryal S and Agarwalla B~K 2021 {\em arXiv\/} (\textit{Preprint}
  \eprint{2106.12809}) \urlprefix\url{https://arxiv.org/abs/2106.12809v2}

\bibitem{scully_quantum_2011}
Scully M~O, Chapin K~R, Dorfman K~E, Kim M~B and Svidzinsky A 2011 {\em Proc.
  Natl. Acad. Sci. U.S.A.\/} {\bf 108} 15097--15100 ISSN 0027-8424
  \urlprefix\url{https://doi.org/10.1073/pnas.1110234108}

\bibitem{brunner_virtual_2012}
Brunner N, Linden N, Popescu S and Skrzypczyk P 2012 {\em Phys. Rev. E\/} {\bf
  85}(5) 051117
  \urlprefix\url{https://link.aps.org/doi/10.1103/PhysRevE.85.051117}

\bibitem{silva_small_2015}
Silva R, Skrzypczyk P and Brunner N 2015 {\em Phys. Rev. E\/} {\bf 92}(1)
  012136 \urlprefix\url{https://link.aps.org/doi/10.1103/PhysRevE.92.012136}

\bibitem{mitchison:2016}
Mitchison M~T, Huber M, Prior J, Woods M~P and Plenio M~B 2016 {\em Quantum
  Sci. Technol.\/} {\bf 1} 015001
  \urlprefix\url{http://stacks.iop.org/2058-9565/1/i=1/a=015001}

\bibitem{hewgill_three_2020}
Hewgill A, Gonz\'alez J~O, Palao J~P, Alonso D, Ferraro A and De~Chiara G 2020
  {\em Phys. Rev. E\/} {\bf 101}(1) 012109
  \urlprefix\url{https://link.aps.org/doi/10.1103/PhysRevE.101.012109}

\bibitem{naseem_two_2020}
Naseem M~T, Misra A and Özgür E~Müstecapl{\i}o{\u{g}}lu 2020 {\em Quantum
  Science and Technology\/} {\bf 5} 035006
  \urlprefix\url{https://doi.org/10.1088/2058-9565/ab8d89}

\bibitem{bhandari_minimal_2021}
Bhandari B and Jordan A~N 2021 {\em Phys. Rev. B\/} {\bf 104}(7) 075442
  \urlprefix\url{https://link.aps.org/doi/10.1103/PhysRevB.104.075442}

\bibitem{poulsen_nonmarkovian_2021}
Poulsen K, Majland M, Lloyd S, Kjaergaard M and Zinner N~T 2021 {\em arXiv\/}
  (\textit{Preprint} \eprint{2108.08855})
  \urlprefix\url{https://arxiv.org/abs/2108.08855v1}

\bibitem{blok_quantum_2021}
Blok M~S, Ramasesh V~V, Schuster T, O{'}Brien K, Kreikebaum J~M, Dahlen D,
  Morvan A, Yoshida B, Yao N~Y and Siddiqi I 2021 {\em Phys. Rev. X\/} {\bf 11}
  021010 ISSN 2160-3308
  \urlprefix\url{https://link.aps.org/doi/10.1103/PhysRevX.11.021010}

\bibitem{brask:2015njp}
Brask J~B, Haack G, Brunner N and Huber M 2015 {\em New J. Phys.\/} {\bf 17}
  113029 \urlprefix\url{http://stacks.iop.org/1367-2630/17/i=11/a=113029}

\bibitem{nesterov_proposal_2021}
Nesterov K~N, Ficheux Q, Manucharyan V~E and Vavilov M~G 2021 {\em PRX
  Quantum\/} {\bf 2}(2) 020345
  \urlprefix\url{https://link.aps.org/doi/10.1103/PRXQuantum.2.020345}

\bibitem{li_quantum_2017}
Li S~W, Kim M~B, Agarwal G~S and Scully M~O 2017 {\em Phys. Rev. A\/} {\bf
  96}(6) 063806
  \urlprefix\url{https://link.aps.org/doi/10.1103/PhysRevA.96.063806}

\bibitem{kalaee_violating_2021}
Kalaee A~A~S, Wacker A and Potts P~P 2021 {\em Phys. Rev. E\/} {\bf 104}(1)
  L012103 \urlprefix\url{https://link.aps.org/doi/10.1103/PhysRevE.104.L012103}

\bibitem{li_negative_2006}
Li B, Wang L and Casati G 2006 {\em Appl. Phys. Lett.\/} {\bf 88} 143501 ISSN
  0003-6951 \urlprefix\url{https://doi.org/10.1063/1.2191730}

\bibitem{wang_thermal_2007}
Wang L and Li B 2007 {\em Phys. Rev. Lett.\/} {\bf 99}(17) 177208
  \urlprefix\url{https://link.aps.org/doi/10.1103/PhysRevLett.99.177208}

\bibitem{li_colloquium_2012}
Li N, Ren J, Wang L, Zhang G, H\"anggi P and Li B 2012 {\em Rev. Mod. Phys.\/}
  {\bf 84}(3) 1045--1066
  \urlprefix\url{https://link.aps.org/doi/10.1103/RevModPhys.84.1045}

\bibitem{segal_spin_2005}
Segal D and Nitzan A 2005 {\em Phys. Rev. Lett.\/} {\bf 94}(3) 034301
  \urlprefix\url{https://link.aps.org/doi/10.1103/PhysRevLett.94.034301}

\bibitem{segal_single_2008}
Segal D 2008 {\em Phys. Rev. Lett.\/} {\bf 100}(10) 105901
  \urlprefix\url{https://link.aps.org/doi/10.1103/PhysRevLett.100.105901}

\bibitem{ojanen_selection_2009}
Ojanen T 2009 {\em Phys. Rev. B\/} {\bf 80}(18) 180301
  \urlprefix\url{https://link.aps.org/doi/10.1103/PhysRevB.80.180301}

\bibitem{ruokola_thermal_2009}
Ruokola T, Ojanen T and Jauho A~P 2009 {\em Phys. Rev. B\/} {\bf 79}(14) 144306
  \urlprefix\url{https://link.aps.org/doi/10.1103/PhysRevB.79.144306}

\bibitem{ruokola:2011}
Ruokola T and Ojanen T 2011 {\em Phys. Rev. B\/} {\bf 83}(24) 241404
  \urlprefix\url{https://link.aps.org/doi/10.1103/PhysRevB.83.241404}

\bibitem{schaller_collective_2016}
Schaller G, Giusteri G~G and Celardo G~L 2016 {\em Phys. Rev. E\/} {\bf 94}(3)
  032135 \urlprefix\url{https://link.aps.org/doi/10.1103/PhysRevE.94.032135}

\bibitem{man_controlling_2016}
Man Z~X, An N~B and Xia Y~J 2016 {\em Phys. Rev. E\/} {\bf 94}(4) 042135
  \urlprefix\url{https://link.aps.org/doi/10.1103/PhysRevE.94.042135}

\bibitem{ordonez-miranda_quantum_2017}
Ordonez-Miranda J, Ezzahri Y and Joulain K 2017 {\em Phys. Rev. E\/} {\bf
  95}(2) 022128
  \urlprefix\url{https://link.aps.org/doi/10.1103/PhysRevE.95.022128}

\bibitem{barzanjeh_manipulating_2018}
Barzanjeh S, Aquilina M and Xuereb A 2018 {\em Phys. Rev. Lett.\/} {\bf 120}(6)
  060601
  \urlprefix\url{https://link.aps.org/doi/10.1103/PhysRevLett.120.060601}

\bibitem{kargi_two_2019}
Karg{\ifmmode \imath \else \i \fi{}} C, Naseem M~T, Opatrn\'y T~c~v,
  M\"ustecapl\ifmmode \imath \else \i \fi{}o\ifmmode~\breve{g}\else
  \u{g}\fi{}lu O~E and Kurizki G 2019 {\em Phys. Rev. E\/} {\bf 99}(4) 042121
  \urlprefix\url{https://link.aps.org/doi/10.1103/PhysRevE.99.042121}

\bibitem{bhandari_thermal_2021}
Bhandari B, Erdman P~A, Fazio R, Paladino E and Taddei F 2021 {\em Phys. Rev.
  B\/} {\bf 103}(15) 155434
  \urlprefix\url{https://link.aps.org/doi/10.1103/PhysRevB.103.155434}

\bibitem{xu_heat_2021}
Xu M, Stockburger J~T and Ankerhold J 2021 {\em Phys. Rev. B\/} {\bf 103}(10)
  104304 \urlprefix\url{https://link.aps.org/doi/10.1103/PhysRevB.103.104304}

\bibitem{poulsen_entanglement_2021}
Poulsen K, Santos A~C, Kristensen L~B and Zinner N~T 2021 {\em arXiv\/}
  (\textit{Preprint} \eprint{2101.04124})
  \urlprefix\url{https://arxiv.org/abs/2101.04124v1}

\bibitem{iorio_photonic_2021}
Iorio A, Strambini E, Haack G, Campisi M and Giazotto F 2021 {\em Phys. Rev.
  Applied\/} {\bf 15}(5) 054050
  \urlprefix\url{https://link.aps.org/doi/10.1103/PhysRevApplied.15.054050}

\bibitem{segal_nonlinear_2008}
Segal D 2008 {\em Phys. Rev. E\/} {\bf 77}(2) 021103
  \urlprefix\url{https://link.aps.org/doi/10.1103/PhysRevE.77.021103}

\bibitem{joulain:2016}
Joulain K, Drevillon J, Ezzahri Y and Ordonez-Miranda J 2016 {\em Phys. Rev.
  Lett.\/} {\bf 116}(20) 200601
  \urlprefix\url{https://link.aps.org/doi/10.1103/PhysRevLett.116.200601}

\bibitem{zhang_coulomb_2018}
Zhang Y, Yang Z, Zhang X, Lin B, Lin G and Chen J 2018 {\em EPL\/} {\bf 122}
  17002 ISSN 0295-5075
  \urlprefix\url{https://doi.org/10.1209/0295-5075/122/17002}

\bibitem{guo_quantum_2018}
Guo B~q, Liu T and Yu C~s 2018 {\em Phys. Rev. E\/} {\bf 98}(2) 022118
  \urlprefix\url{https://link.aps.org/doi/10.1103/PhysRevE.98.022118}

\bibitem{majland_quantum_2020}
Majland M, Christensen K~S and Zinner N~T 2020 {\em Phys. Rev. B\/} {\bf
  101}(18) 184510
  \urlprefix\url{https://link.aps.org/doi/10.1103/PhysRevB.101.184510}

\bibitem{tahir_minimal_2020}
Naseem M~T, Misra A, M\"ustecaplio\ifmmode~\breve{g}\else \u{g}\fi{}lu O~E and
  Kurizki G 2020 {\em Phys. Rev. Research\/} {\bf 2}(3) 033285
  \urlprefix\url{https://link.aps.org/doi/10.1103/PhysRevResearch.2.033285}

\bibitem{entin:2010}
Entin-Wohlman O, Imry Y and Aharony A 2010 {\em Phys. Rev. B\/} {\bf 82}(11)
  115314 \urlprefix\url{http://link.aps.org/doi/10.1103/PhysRevB.82.115314}

\bibitem{hotspots}
S\'anchez R and B\"uttiker M 2011 {\em Phys. Rev. B\/} {\bf 83}(8) 085428
  \urlprefix\url{http://link.aps.org/doi/10.1103/PhysRevB.83.085428}

\bibitem{thierschmann:2015}
Thierschmann H, S\'anchez R, Sothmann B, Arnold F, Heyn C, Hansen W, Buhmann H
  and Molenkamp L~W 2015 {\em Nat. Nanotechnol.\/} {\bf 10} 854 ISSN 1748-3387
  \urlprefix\url{http://dx.doi.org/10.1038/nnano.2015.176}

\bibitem{koski:2015}
Koski J~V, Kutvonen A, Khaymovich I~M, Ala-Nissila T and Pekola J~P 2015 {\em
  Phys. Rev. Lett.\/} {\bf 115}(26) 260602
  \urlprefix\url{https://link.aps.org/doi/10.1103/PhysRevLett.115.260602}

\bibitem{dorsch:2020}
Dorsch S, Svilans A, Josefsson M, Goldozian B, Kumar M, Thelander C, Wacker A
  and Burke A 2021 {\em Nano Lett.\/} {\bf 21} 988--994 ISSN 1530-6984
  \urlprefix\url{https://doi.org/10.1021/acs.nanolett.0c04017}

\bibitem{maslennikov:2019}
Maslennikov G, Ding S, Hablutzel R, Gan J, Roulet A, Nimmrichter S, Dai J,
  Scarani V and Matsukevich D 2019 {\em Nat. Commun.\/} {\bf 10} 202
  \urlprefix\url{http://doi.org/10.1038/s41467-018-08090-0}

\bibitem{vool_introduction_2017}
Vool U and Devoret M 2017 {\em Int. J. Circuit Theory Appl.\/} {\bf 45}
  897--934 ISSN 0098-9886 \urlprefix\url{https://doi.org/10.1002/cta.2359}

\bibitem{blais_circuit_2021}
Blais A, Grimsmo A~L, Girvin S~M and Wallraff A 2021 {\em Rev. Mod. Phys.\/}
  {\bf 93}(2) 025005
  \urlprefix\url{https://link.aps.org/doi/10.1103/RevModPhys.93.025005}

\bibitem{krantz_quantum_2019}
Krantz P, Kjaergaard M, Yan F, Orlando T~P, Gustavsson S and Oliver W~D 2019
  {\em Appl. Phys. Rev.\/} {\bf 6} 021318 ISSN 1931-9401
  \urlprefix\url{https://doi.org/10.1063/1.5089550}

\bibitem{baur_measurement_2009}
Baur M, Filipp S, Bianchetti R, Fink J~M, G\"oppl M, Steffen L, Leek P~J, Blais
  A and Wallraff A 2009 {\em Phys. Rev. Lett.\/} {\bf 102}(24) 243602
  \urlprefix\url{https://link.aps.org/doi/10.1103/PhysRevLett.102.243602}

\bibitem{sillanpaa_autler_2009}
Sillanp\"a\"a M~A, Li J, Cicak K, Altomare F, Park J~I, Simmonds R~W, Paraoanu
  G~S and Hakonen P~J 2009 {\em Phys. Rev. Lett.\/} {\bf 103}(19) 193601
  \urlprefix\url{https://link.aps.org/doi/10.1103/PhysRevLett.103.193601}

\bibitem{bianchetti_control_2010}
Bianchetti R, Filipp S, Baur M, Fink J~M, Lang C, Steffen L, Boissonneault M,
  Blais A and Wallraff A 2010 {\em Phys. Rev. Lett.\/} {\bf 105}(22) 223601
  \urlprefix\url{https://link.aps.org/doi/10.1103/PhysRevLett.105.223601}

\bibitem{kelly_direct_2010}
Kelly W~R, Dutton Z, Schlafer J, Mookerji B, Ohki T~A, Kline J~S and Pappas D~P
  2010 {\em Phys. Rev. Lett.\/} {\bf 104}(16) 163601
  \urlprefix\url{https://link.aps.org/doi/10.1103/PhysRevLett.104.163601}

\bibitem{abdumalikov_electromagnetically_2010}
Abdumalikov A~A, Astafiev O, Zagoskin A~M, Pashkin Y~A, Nakamura Y and Tsai J~S
  2010 {\em Phys. Rev. Lett.\/} {\bf 104}(19) 193601
  \urlprefix\url{https://link.aps.org/doi/10.1103/PhysRevLett.104.193601}

\bibitem{honiglDecrinis_mixing_2018}
H\"onigl-Decrinis T, Antonov I~V, Shaikhaidarov R, Antonov V~N, Dmitriev A~Y
  and Astafiev O~V 2018 {\em Phys. Rev. A\/} {\bf 98}(4) 041801
  \urlprefix\url{https://link.aps.org/doi/10.1103/PhysRevA.98.041801}

\bibitem{tan_topological_2018}
Tan X, Zhang D~W, Liu Q, Xue G, Yu H~F, Zhu Y~Q, Yan H, Zhu S~L and Yu Y 2018
  {\em Phys. Rev. Lett.\/} {\bf 120}(13) 130503
  \urlprefix\url{https://link.aps.org/doi/10.1103/PhysRevLett.120.130503}

\bibitem{vepsalainen_simulating_2020}
Veps{\ifmmode\ddot{a}\else\"{a}\fi}l{\ifmmode\ddot{a}\else\"{a}\fi}inen A and
  Paraoanu G~S 2020 {\em Adv. Quantum Technol.\/} {\bf 3} 1900121 ISSN
  2511-9044 \urlprefix\url{https://doi.org/10.1002/qute.201900121}

\bibitem{fedorov_implementation_2012}
Fedorov A, Steffen L, Baur M, da~Silva M~P and Wallraff A 2012 {\em Nature\/}
  {\bf 481} 170--172 ISSN 1476-4687
  \urlprefix\url{hrrps://doi.org/10.1038/nature10713}

\bibitem{abdumalikov_experimental_2013}
Abdumalikov~A Jr A, Fink J~M, Juliusson K, Pechal M, Berger S, Wallraff A and
  Filipp S 2013 {\em Nature\/} {\bf 496} 482--485 ISSN 1476-4687
  \urlprefix\url{https://doi.org/10.1038/nature12010}

\bibitem{yurtalan_implementation_2020}
Yurtalan M~A, Shi J, Kononenko M, Lupascu A and Ashhab S 2020 {\em Phys. Rev.
  Lett.\/} {\bf 125}(18) 180504
  \urlprefix\url{https://link.aps.org/doi/10.1103/PhysRevLett.125.180504}

\bibitem{morvan_qutrit_2021}
Morvan A, Ramasesh V~V, Blok M~S, Kreikebaum J~M, O'Brien K, Chen L, Mitchell
  B~K, Naik R~K, Santiago D~I and Siddiqi I 2021 {\em Phys. Rev. Lett.\/} {\bf
  126}(21) 210504
  \urlprefix\url{https://link.aps.org/doi/10.1103/PhysRevLett.126.210504}

\bibitem{cervera-Lierta_experimental_2021}
Cervera-Lierta A, Krenn M, Aspuru-Guzik A and Galda A 2021 {\em arXiv\/}
  (\textit{Preprint} \eprint{2104.05627})
  \urlprefix\url{https://arxiv.org/abs/2104.05627v2}

\bibitem{jerger_contextuality_2016}
Jerger M, Reshitnyk Y, Oppliger M, Poto{\ifmmode\check{c}\else\v{c}\fi}nik A,
  Mondal M, Wallraff A, Goodenough K, Wehner S, Juliusson K, Langford N~K and
  Fedorov A 2016 {\em Nat. Commun.\/} {\bf 7} 1--6 ISSN 2041-1723
  \urlprefix\url{https://doi.org/10.1038/ncomms12930}

\bibitem{palacios-Laloy_tunable_2008}
Palacios-Laloy A, Nguyen F, Mallet F, Bertet P, Vion D and Esteve D 2008 {\em
  J. Low Temp. Phys.\/} {\bf 151} 1034--1042 ISSN 1573-7357
  \urlprefix\url{https://doi.org/10.1007/s10909-008-9774-x}

\bibitem{goppl_coplanar_2008}
G{\ifmmode\ddot{o}\else\"{o}\fi}ppl M, Fragner A, Baur M, Bianchetti R, Filipp
  S, Fink J~M, Leek P~J, Puebla G, Steffen L and Wallraff A 2008 {\em J. Appl.
  Phys.\/} {\bf 104} 113904 ISSN 0021-8979
  \urlprefix\url{https://doi.org/10.1063/1.3010859}

\bibitem{partanen_quantum_2016}
Partanen M, Tan K~Y, Govenius J, Lake R~E,
  M{\ifmmode\ddot{a}\else\"{a}\fi}kel{\ifmmode\ddot{a}\else\"{a}\fi} M~K,
  Tanttu T and
  M{\ifmmode\ddot{o}\else\"{o}\fi}tt{\ifmmode\ddot{o}\else\"{o}\fi}nen M 2016
  {\em Nat. Phys.\/} {\bf 12} 460--464 ISSN 1745-2481
  \urlprefix\url{https://doi.org/10.1038/nphys3642}

\bibitem{giazotto:2006}
Giazotto F, Heikkil\"a T~T, Luukanen A, Savin A~M and Pekola J~P 2006 {\em Rev.
  Mod. Phys.\/} {\bf 78}(1) 217
  \urlprefix\url{http://link.aps.org/doi/10.1103/RevModPhys.78.217}

\bibitem{pekola_colloquium_2021}
Pekola J~P and Karimi B 2021 {\em Rev. Mod. Phys.\/} {\bf 93}(4) 041001
  \urlprefix\url{https://link.aps.org/doi/10.1103/RevModPhys.93.041001}

\bibitem{chen_quantum_2012}
Chen Y~X and Li S~W 2012 {\em EPL\/} {\bf 97} 40003
  \urlprefix\url{https://doi.org/10.1209/0295-5075/97/40003}

\bibitem{hofer_autonomous_2016}
Hofer P~P, Perarnau-Llobet M, Brask J~B, Silva R, Huber M and Brunner N 2016
  {\em Phys. Rev. B\/} {\bf 94}(23) 235420
  \urlprefix\url{http://10.1103/PhysRevB.94.235420}

\bibitem{hofer_quantum_2016}
Hofer P~P, Souquet J~R and Clerk A~A 2016 {\em Phys. Rev. B\/} {\bf 93}(4)
  041418(R) \urlprefix\url{https://10.1103/PhysRevB.93.041418}

\bibitem{thomas_thermally_2020}
Thomas G, Gubaydullin A, Golubev D~S and Pekola J~P 2020 {\em Phys. Rev. B\/}
  {\bf 102}(10) 104503
  \urlprefix\url{https://link.aps.org/doi/10.1103/PhysRevB.102.104503}

\bibitem{karimi_coupled_2017}
Karimi B, Pekola J~P, Campisi M and Fazio R 2017 {\em Quantum Sci. Technol.\/}
  {\bf 2} 044007 ISSN 2058-9565
  \urlprefix\url{https://doi.org/10.1088/2058-9565/aa8330}

\bibitem{ronzani_tunable_2018}
Ronzani A, Karimi B, Senior J, Chang Y~C, Peltonen J~T, Chen C and Pekola J~P
  2018 {\em Nat. Phys.\/} {\bf 14} 991--995 ISSN 1745-2481
  \urlprefix\url{https://doi.org/10.1038/s41567-018-0199-4}

\bibitem{senior_heat_2020}
Senior J, Gubaydullin A, Karimi B, Peltonen J~T, Ankerhold J and Pekola J~P
  2020 {\em Commun. Phys.\/} {\bf 3} 1--5 ISSN 2399-3650
  \urlprefix\url{https://doi.org/10.1038/s42005-020-0307-5}

\bibitem{gubaydullin_photonic_2021}
Gubaydullin A, Thomas G, Golubev D~S, Lvov D, Peltonen J~T and Pekola J~P 2021
  {\em arXiv\/} (\textit{Preprint} \eprint{2112.09224})
  \urlprefix\url{https://arxiv.org/abs/2112.09224v1}

\bibitem{sanchez_single_2017}
S{\ifmmode\acute{a}\else\'{a}\fi}nchez R, Thierschmann H and Molenkamp L~W 2017
  {\em New J. Phys.\/} {\bf 19} 113040 ISSN 1367-2630
  \urlprefix\url{https://doi.org/10.1088/1367-2630/aa8b94}

\bibitem{strasberg:2018}
Strasberg P, Schaller G, Schmidt T~L and Esposito M 2018 {\em Phys. Rev. B\/}
  {\bf 97}(20) 205405
  \urlprefix\url{https://link.aps.org/doi/10.1103/PhysRevB.97.205405}

\bibitem{hwang:2018}
Hwang S~Y, Giazotto F and Sothmann B 2018 {\em Phys. Rev. Appl.\/} {\bf 10}
  044062 ISSN 2331-7019
  \urlprefix\url{https://doi.org/10.1103/PhysRevApplied.10.044062}

\bibitem{acciai_phase_2021}
Acciai M, Hajiloo F, Hassler F and Splettstoesser J 2021 {\em Phys. Rev. B\/}
  {\bf 103}(8) 085409
  \urlprefix\url{https://link.aps.org/doi/10.1103/PhysRevB.103.085409}

\bibitem{breuer:book}
Breuer H~P and Petruccione F 2002 {\em The theory of open quantum systems\/}
  (Oxford University Press)
  \urlprefix\url{https://doi.org/10.1093/acprof:oso/9780199213900.001.0001}

\bibitem{gernot}
Schaller G 2014 {\em {Open Quantum Systems Far from Equilibrium}\/} (Cham,
  Switzerland: Springer)
  \urlprefix\url{https://doi.org/10.1007/978-3-319-03877-3}

\bibitem{clarke_superconducting_2008}
Clarke J and Wilhelm F~K 2008 {\em Nature\/} {\bf 453} 1031--1042 ISSN
  1476-4687 \urlprefix\url{https://doi.org/10.1038/nature07128}

\bibitem{manzano_hybrid_2020}
Manzano G, S\'anchez R, Silva R, Haack G, Brask J~B, Brunner N and Potts P~P
  2020 {\em Phys. Rev. Research\/} {\bf 2}(4) 043302
  \urlprefix\url{https://link.aps.org/doi/10.1103/PhysRevResearch.2.043302}

\bibitem{humphrey:02}
Humphrey T~E, Newbury R, Taylor R~P and Linke H 2002 {\em Phys. Rev. Lett.\/}
  {\bf 89}(11) 116801
  \urlprefix\url{https://link.aps.org/doi/10.1103/PhysRevLett.89.116801}

\bibitem{nakpathomkun:2010}
Nakpathomkun N, Xu H~Q and Linke H 2010 {\em Phys. Rev. B\/} {\bf 82} 235428
  ISSN 2469-9969 \urlprefix\url{https://doi.org/10.1103/PhysRevB.82.235428}

\bibitem{jordan:2013}
Jordan A~N, Sothmann B, S\'anchez R and B\"uttiker M 2013 {\em Phys. Rev. B\/}
  {\bf 87}(7) 075312
  \urlprefix\url{http://link.aps.org/doi/10.1103/PhysRevB.87.075312}

\bibitem{benenti_from_2016}
Benenti G, Casati G, Mej{\ifmmode\acute{\imath}\else\'{\i}\fi}a-Monasterio C
  and Peyrard M 2016 {From Thermal Rectifiers to Thermoelectric Devices} {\em
  {Thermal Transport in Low Dimensions}\/} (Cham, Switzerland: Springer) pp
  365--407 ISBN 978-3-319-29259-5
  \urlprefix\url{https://doi.org/10.1007/978-3-319-29261-8_10}

\bibitem{martinez-Perez_rectification_2015}
Mart{\ifmmode\acute{\imath}\else\'{\i}\fi}nez-P{\ifmmode\acute{e}\else\'{e}\fi}rez
  M~J, Fornieri A and Giazotto F 2015 {\em Nat. Nanotechnol.\/} {\bf 10}
  303--307 ISSN 1748-3395 \urlprefix\url{https://doi.org/10.1038/nnano.2015.11}

\bibitem{jiang:2015}
Jiang J~H, Kulkarni M, Segal D and Imry Y 2015 {\em Phys. Rev. B\/} {\bf 92}
  045309 ISSN 2469-9969
  \urlprefix\url{https://doi.org/10.1103/PhysRevB.92.045309}

\bibitem{donald}
Goury D and S{\ifmmode\acute{a}\else\'{a}\fi}nchez R 2019 {\em Appl. Phys.
  Lett.\/} {\bf 115} 092601 ISSN 0003-6951
  \urlprefix\url{https://doi.org/10.1063/1.5109100}

\bibitem{rossello:2017}
Rossell{\ifmmode\acute{o}\else\'{o}\fi} G, L{\ifmmode\acute{o}\else\'{o}\fi}pez
  R and S{\ifmmode\acute{a}\else\'{a}\fi}nchez R 2017 {\em Phys. Rev. B\/} {\bf
  95} 235404 ISSN 2469-9969
  \urlprefix\url{https://doi.org/10.1103/PhysRevB.95.235404}

\bibitem{chiraldiode}
S{\ifmmode\acute{a}\else\'{a}\fi}nchez R, Sothmann B and Jordan A~N 2015 {\em
  New J. Phys.\/} {\bf 17} 075006 ISSN 1367-2630
  \urlprefix\url{https://doi.org/10.1088/1367-2630/17/7/075006}

\bibitem{zhang_ballistic_2010}
Zhang L, Wang J~S and Li B 2010 {\em Phys. Rev. B\/} {\bf 81}(10) 100301
  \urlprefix\url{https://link.aps.org/doi/10.1103/PhysRevB.81.100301}

\bibitem{ming_ballistic_2010}
Ming Y, Wang Z~X, Ding Z~J and Li H~M 2010 {\em New Journal of Physics\/} {\bf
  12} 103041 \urlprefix\url{https://doi.org/10.1088/1367-2630/12/10/103041}

\bibitem{alex}
Marcos-Vicioso A, L\'opez-Jurado C, Ruiz-Garcia M and S\'anchez R 2018 {\em
  Phys. Rev. B\/} {\bf 98}(3) 035414
  \urlprefix\url{https://link.aps.org/doi/10.1103/PhysRevB.98.035414}

\bibitem{gelbwaserKlimovsky_work_2015}
Gelbwaser-Klimovsky D and Kurizki G 2015 {\em Sci. Rep.\/} {\bf 5} 1--6 ISSN
  2045-2322 \urlprefix\url{https://doi.org/10.1038/srep07809}

\end{thebibliography}

\end{document}